\documentclass[journal]{IEEEtran}
\usepackage{graphicx}
\usepackage{amsmath}
\usepackage{amssymb}
\usepackage{tabu}
\usepackage{cite}
\usepackage{indentfirst}
\usepackage[utf8]{inputenc}
\usepackage[english]{babel}
\usepackage{bbm} 
\usepackage{epstopdf}
\usepackage{amsxtra}
\usepackage{amstext}
\usepackage{amsthm}
\usepackage{latexsym}
\usepackage{dsfont} 
\usepackage{xcolor}
\usepackage{units}
\usepackage{multirow}
\usepackage{lipsum}
\usepackage{bm}
\usepackage{multicol}
\usepackage{setspace}
\usepackage{caption}
\usepackage{subcaption}
\usepackage{soul}
\usepackage[bottom]{footmisc}
\usepackage{tikz}

\usetikzlibrary{arrows}
\usepackage[inline, shortlabels]{enumitem}

\definecolor{darkgreen}{rgb}{0,0.75,0}
\definecolor{maroon}{RGB}{186,0,0}
\definecolor{purple}{RGB}{96,26,149}
\definecolor{mavi}{RGB}{46,76,255}
\definecolor{haki}{RGB}{38,99,33}

\makeatletter
\let\NAT@parse\undefined
\makeatother

\usepackage{acro}
\DeclareAcronym{RSSI}{
  short = RSSI,
  long  = received signal strength indicator
}

\DeclareAcronym{6G}{
  short = 6G,
  long  = sixth-generation
}
\DeclareAcronym{RSS}{
  short = RSS,
  long  = received signal strength
}
\DeclareAcronym{ED}{
  short = ED,
  long  = enhanced distance
}
\DeclareAcronym{OFDM}{
  short = OFDM,
  long  = orthogonal frequency division multiplexing
}
\DeclareAcronym{TTD}{
  short = TTD,
  long  = true-time-delay
}
\DeclareAcronym{ULA}{
  short = ULA,
  long  = uniform linear array
}
\DeclareAcronym{UPA}{
  short = UPA,
  long  = uniform planar array
}
\DeclareAcronym{UCA}{
  short = UCA,
  long  = uniform circular array
}
\DeclareAcronym{LoS}{
  short = LoS,
  long  = line of sight
}
\DeclareAcronym{CFO}{
  short = CFO,
  long  = carrier frequency offset
}
\DeclareAcronym{TDD}{
  short = TDD,
  long  = time division duplexing
}
\DeclareAcronym{FDD}{
  short = FDD,
  long  = frequency division duplexing
}
\DeclareAcronym{EKF}{
  short = EKF,
  long  = extended Kalman filter
}
\DeclareAcronym{UWB}{
  short = UWB,
  long  = ultra wideband
}
\DeclareAcronym{AWGN}{
  short = AWGN,
  long  = additive white Gaussian noise
}
\DeclareAcronym{MIMO}{
  short = MIMO,
  long  = multiple-input multiple-output
}
\DeclareAcronym{AoA}{
  short = AoA ,
  long  = angle of arrival
}
\DeclareAcronym{AoD}{
  short = AoD,
  long  = angle of departure
}
\DeclareAcronym{SNR}{
  short = SNR,
  long  = signal-to-noise ratio
}
\DeclareAcronym{BER}{
  short = BER,
  long  = bit error rate
}
\DeclareAcronym{DFT}{
  short = DFT,
  long  = discrete Fourier transform
}
\DeclareAcronym{3GPP}{
  short = 3GPP,
  long  = third generation partnership project
}
\DeclareAcronym{BS}{
  short = BS,
  long  = base station
}
\DeclareAcronym{AP}{
  short = AP,
  long  = access point
}
\DeclareAcronym{PL}{
  short = PL,
  long  = physical layer
}
\DeclareAcronym{PLA}{
  short = PLA,
  long  = physical layer authentication
}
\DeclareAcronym{mmWave}{
  short = mmWave,
  long  = millimeter wave
}
\DeclareAcronym{CIR}{
  short = CIR,
  long  = channel impulse response
}
\DeclareAcronym{RF}{
  short = RF,
  long  = radio frequency 
}
\DeclareAcronym{ROC}{
  short = ROC,
  long  = receiver operating characteristics 
}
 \DeclareAcronym{RIS}{
  short = RIS,
  long  = Reconfigurable Intelligent Surfaces 
}

\DeclareAcronym{PLS}{
  short = PLS,
  long  = Physical Layer Security 
}

\DeclareAcronym{THz}{
  short =  THz,
  long  = Terahertz
}

\theoremstyle{remark}

\theoremstyle{definition}
\usepackage{algorithm}
\usepackage{algpseudocode}
\algnewcommand{\LineComment}[1]{\State \(\triangleright\) #1}
\newcommand{\nth}[1]{{#1}^{\text{th}}}
\newcommand{\mbf}[1]{\mathbf{#1}}

\newcommand{\expec}[1]{{\mathbb{E}\!\left[{#1}\right]}}

\newcommand{\NBS}[0]{N_{\mathrm{BS}}}
\newcommand{\OBS}[0]{O_{\mathrm{BS}}}
\newcommand{\MBS}[0]{M_{\mathrm{BS}}}
\newcommand{\NRF}[0]{N_{\mathrm{RF}}}

\newcommand{\FBB}[0]{\mathbf{F}^{\mathrm{BB}}}
\newcommand{\FRF}[0]{\mathbf{F}^{\mathrm{RF}}}
\begin{document}
\title{
Explainable E2E Learning for Probing Beams and RSSI-Based Multi-User Hybrid Precoding Design
}
\title{Explainable Autoencoder Design for RSSI-Based Multi-User Beam Probing and Hybrid Precoding}
	\author
	{
		{  Asmaa~Abdallah,~\IEEEmembership{Member,~IEEE}, Abdulkadir~Celik,~\IEEEmembership{Senior~Member,~IEEE},\\
		Ahmed~Alkhateeb,~\IEEEmembership{Member,~IEEE},
		and Ahmed~M.~Eltawil,~\IEEEmembership{Senior Member,~IEEE}. 	}	

		\thanks{\scriptsize A. Abdallah, A. Celik, and A. M. Eltawil are with Computer, Electrical, and Mathematical Sciences and Engineering (CEMSE) Division, King Abdullah University of Science and Technology (KAUST), Thuwal, 23955-6900, KSA.\\ A. Alkhateeb is with with the Wireless Intelligence Lab at the School of Electrical, Computer and Energy Engineering, Arizona State University (ASU).\\ The authors gratefully acknowledge financial support for this work from KAUST.}}
\maketitle
	
	

\begin{abstract}
This paper introduces a novel neural network (NN) structure referred to as an ``Auto-hybrid precoder'' (Auto-HP) and an unsupervised deep learning (DL) approach that jointly designs \ac{mmWave} probing beams and hybrid precoding matrix design for mmWave multi-user communication system  with minimal training pilots. Our learning-based model capitalizes on prior channel observations to achieve two primary goals: designing a limited set of probing beams and predicting off-grid \ac{RF} beamforming vectors. The Auto-HP framework optimizes the probing beams in an unsupervised manner, concentrating the sensing power on the most promising spatial directions based on the surrounding environment. This is achieved through an innovative neural network architecture that respects \ac{RF} chain constraints and models received signal strength power measurements using complex-valued convolutional layers. Then, the autoencoder is trained to directly produce RF beamforming vectors for hybrid architectures, unconstrained by a predefined codebook, based on few projected received signal strength indicators (RSSIs). Finally, once the RF beamforming vectors for the multi-users are predicted, the baseband (BB) digital precoders are designed accounting for the multi-user interference. The Auto-HP neural network is trained end-to-end (E2E) in an unsupervised learning  manner with a customized loss function that aims to maximizes the received signal strength. 
The adequacy of the Auto-HP NN's bottleneck layer dimension is evaluated from an information theory perspective, ensuring maximum data compression and reliable RF beam predictions.  The proposed algorithm leverages mutual information (MI) and entropy evolution during Auto-HP training, reducing training overhead significantly. In a system with 64 antennas, 4 RF chains, and 4 users, our solution requires only 8 training pilots to predict RF beamforming/combining vectors directly, and achieve near-optimal data rates.
\end{abstract}
	

\section{Introduction}\label{sec:Intro}
Millimeter-wave (mmWave) devices necessitate precise beamforming to address substantial isotropic path loss and ensure sufficient signal strength. Analog beams are commonly utilized in practical mmWave systems to minimize costs and power consumption, directing energy toward specific angles. However, selecting the optimal focused beams is highly sensitive to the variations in the propagation environment, including blockages and reflections. For both base stations (BSs) and user equipments (UEs), it is essential to identify effective beamforming directions during initial connection and subsequently track these beams amidst variations in propagation conditions, such as UE movements and rotations. 
As cellular systems advance to wider bandwidths and higher carrier frequencies, including the "sub-terahertz (THz)" bands reaching up to 300 GHz, devices will incorporate increasingly dense antenna arrays and narrower beams. Consequently, beam management, encompassing the discovery and maintenance of optimal beamforming directions, emerges as a formidable challenge, particularly as we advance towards 6G and beyond.  Currently, there are primarily two types of beamforming techniques: 

\begin{enumerate}
    \item \textbf{Channel state information  (CSI) based beamforming techniques}:   The initial phase of CSI-based beamforming involves the estimation of channel state information, typically conducted through the transmission of pilot signals. Following channel estimation, UEs in the downlink scenario commonly feed back this information back to the BS. 
    Finally, with the knowledge of the channel state, the BS uses complex optimization based beamforming algorithms  to design the beamforming vectors or matrices \cite{G2019WSR,Liu2021BF,Rehman2021BF,Wang2021BF,Icc202BFRIS,2019WSBF,SMRIS2021WCNC}. 

    \item \textbf{Beam sweeping-based techniques}: In massive multiple-input multiple-output (MIMO) systems, acquiring perfect CSI presents challenges owing to limitations in signaling overhead. To overcome this challenge, beamforming codebooks offer a viable solution. During the initial access phase, synchronization bursts contain a subset of beams that the BS typically scans. Subsequently, UEs report back to the BS the index of the beam with the strongest received power. By leveraging beam codebooks, MIMO systems can reduce the overhead associated with channel training while maximizing received signal strengths. It is worth noting that the efficiency of training through beam sweeping is intricately tied to the beam codebook's design.
\end{enumerate}

\begin{figure}
    \centering
    \includegraphics[width=0.9\linewidth]{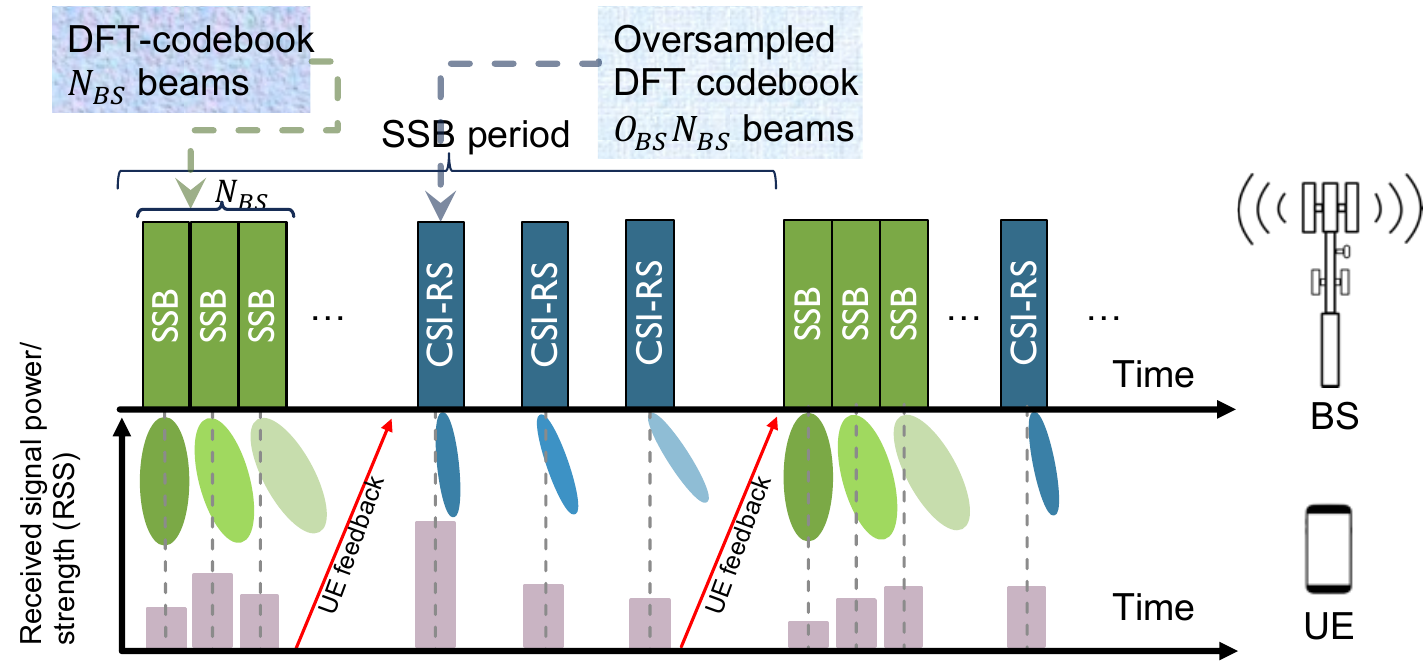}
    \caption{Beam alignment classical procedure, where $\NBS$ is the number of antennas at BS and $\OBS$ is the oversampling factor.}
    \label{fig:beamConv}
\end{figure}

The current implementation of 5G integrates a beam alignment framework that focuses on beam sweeping, measurements, and reporting \cite{Gior2019BM, Li2020BM, Andrew2021BM}. As illustrated in Fig. \ref{fig:beamConv}, in the downlink, the BS employs synchronization signal blocks (SSBs) for sweeping wide probing beams selected from codebooks with quantized beam directions and more flexible CSI reference signals (CSI-RSs) for refining narrow beams. The UE monitors the received signal power, uses either a quasi-omnidirectional beam or sweeps its beam codebook with different receiving beams, and then communicates the received signal measurements back to the BS. Quantized beams generally distribute energy evenly across angular space, commonly employing on-grid codebooks such as discrete Fourier transform (DFT) and oversampled-DFT (O-DFT) codebooks to  ensure coverage across any location. One major issue of the on-grid methods is the ``grid mismatch'', where the predefined beams may not align perfectly with the actual channel conditions, leading to suboptimal performance. 

SSBs are transmitted periodically and serve various purposes, including cell discovery and initial access (IA) for new UEs. To synchronize before connecting to the network, an unconnected UE must measure the SSBs from the Base Station (BS), identify one with a favorable beam, and extract the necessary information.  
The prominence of beam sweeping in both beam alignment and cell discovery suggests that  future releases of 5G will continue to utilize beam sweeping-based frameworks. However, a significant drawback of exhaustive search is the linear increase in beam sweeping latency with the total number of beams and antennas deployed. 
As systems progress to higher frequencies and adopt more antennas with narrower beams, the codebook size increases accordingly, leading to substantial overhead and latency in beam sweeping. In a THz system, the exhaustive search may involve tens of thousands of beam pairs due to the extremely large antenna arrays \cite{daiELAA2023}, rendering it impractical due to the associated prohibitive beam sweeping latency. 

Moreover, while probing beams discover or update channel characteristics by sending pilot signals to gather information on how the channel behaves, precoding beam design utilizes beam probing to enhance system performance by effectively managing the transmission beams towards intended UEs. The BS can determine which directions offer the best signal paths based on the UE feedback by probing the channel with beams directed in multiple orientations. This directional information is key to constructing or updating the precoding matrix, which consists of beamforming vectors tailored to maximize energy toward intended users while nullifying it toward unintended UEs or interferers. 
\subsection{Related Work}
\textit{CSI-Based Beamforming Techniques:}
Unlike  predefined codebook-based  beam management techniques, where the beamforming gain is confined by the granularity of the codebook, a grid-free strategy expands the exploration scope for the optimal beam into a high-dimensional continuous space, theoretically achieving superior gain. For example, singular-value decomposition (SVD) based beamforming, maximum ratio transmission (MRT), and equal gain transmission (EGT) are considered as efficient beamforming techniques under a unit power constraint. However, these methods necessitate complete CSI for computing beamforming weights, a challenge as CSI is typically unavailable before beam alignment. Moreover, even with CSI, determining the optimal SVD and equal gain combining (EGC) beams in a MIMO setup entails a grid search over potential weights, rendering it computationally impractical \cite{Heath2016OverviewmmWave}. Recent research has delved into directly predicting hybrid beamforming weights for the BS without relying on full CSI, leveraging learning techniques applied to sensing matrices \cite{attiah2022DLCSHP} or a series of interactive sensing vectors based on feedback from prior measurements \cite{Sohrabi2022ActiveSens}.

\textit{Codebook-based Beam Selection Approaches:}
Moreover, previous studies in \cite{Andrews2022BF, li2019deep} introduced beam alignment methods based on codebooks, utilizing measurements from specific probing beams at each site to predict the index of the optimal narrow transmit beam chosen from the DFT-based codebook. However, its effectiveness relies significantly on trying multiple candidate beams for each UE. Notably, these studies have only addressed the single-user setup, overlooking multi-user interference, where beam sweeping is employed for each UE. Consequently, the advantage of reducing beam sweeping overhead diminishes as the number of UEs increases. This limitation is evident in various existing methods, such as hierarchical searches that involve exploring all child beams, contextual information (CI) based models \cite{Khateeb2023GPS,khateeb2022radar,Salehi2022, Khateeb20203D}, where the best beams  for a connection is inherently linked to the topology of the propagation environment, as well as the locations of both BS and UE.  
The CI-based models necessitate trials of the most likely candidates, while active learning-based methods require a unique sequence of sensing beams for each UE. Given that each UE may require a distinct set of beams, the beam sweeping process must be reiterated for all UEs within the cell. The overall latency incurred by beam sweeping escalates proportionally with the number of UEs, a factor that can be significant and unpredictable in cellular systems. Ultimately, in codebook-based approaches, exhaustive exploration of the entire codebook may become necessary, presenting a significant limitation in the aforementioned approaches. 

{

\textit{Beam Training Overhead Reduction Mechanisms:} A hierarchical search approach can mitigate latency for a single UE by sweeping broader beams first and progressively transitioning to narrower child beams, iteratively reducing the search space.  However,  the hierarchical search is more susceptible to search errors induced by noisy measurements and imperfect wide-beam shapes \cite{dobre2020Hier,Asmaa2024MADRL}. An alternative strategy to alleviate beam sweeping latency involves utilizing CI. Spatial characteristics, encompassed by CI such as the global positioning system (GPS) coordinates of the UE \cite{Khateeb2023GPS}, radar measurements \cite{khateeb2022radar}, and environmental data derived from images and 3D point clouds \cite{Salehi2022, Khateeb20203D}, can capture such pertinent details. By establishing a correlation from CI to the index of the optimal beam through methods such as a lookup table or data driven models, the search space can be significantly reduced. Nevertheless, these CI-centric approaches face hurdles in standardization and are not widely embraced in practical applications due to the additional sensor requirements and varying hardware capabilities of mmWave devices. Additionally, they might be unsuitable for standalone systems, given the need for a robust feedback link, typically occupying a lower frequency band.

\textit{Beam Prediction and Site-Specific Codebook Design:}
Inspired by the successes of machine learning (ML) and artificial intelligence (AI) schemes, recent research has explored the beam management application of deep learning (DL), wherein deep neural networks (DNN) can extract features from additional information such as CI \cite{Khateeb2023GPS, khateeb2022radar, Salehi2022, Khateeb20203D} and historical measurements \cite{Lim2021BT}, enabling the prediction of the optimal beam. An effective approach to mitigate beam sweeping latency involves tailoring the system to the specific site characteristics. In scenarios with non-uniform angle-of-arrival (AoA) and angle-of-departure (AoD) distributions of channel paths, such as those observed in certain environmental topologies or areas with multiple spatial UE clusters, certain beams in a uniformly distributed codebook may remain unused. Exploiting this observation, the BS can analyze the statistics of the entire codebook following an exhaustive training phase and prioritize the most frequently utilized beams \cite{Liu2019statBM}. Reinforcement learning (RL) approaches have also been proposed to acquire beam management policies through iterative interactions with the environment \cite{DRL2023BM}, training with explicit CSI \cite{khateeb2022BM}, or through the design of site-specific beam codebooks \cite{Asmaa2024MADRL, zhang2021reinforcement, AsmaaCCNC2023MADRL, AsmaaGLOBECOM2023MADRL, heng2023grid}. DNNs can generate small-sized codebooks from scratch, dynamically adapting to the environment and efficiently directing energy toward the UEs. This data-driven approach can also aid in selecting analog beams from standard codebooks through sensing measurements by co-learning the sensing beams and the mapping function \cite{li2019deep}.\footnote{For a comprehensive understanding of beam management solutions using ML and DL, we refer readers to \cite{Hanzo2023BA, khan2023BM}.}
}

Despite DL's impressive performance, the lack of transparency in model decisions hinders its deployment for mission-critical services \cite{XAInetworking}. Establishing quantifiable metrics for DL-based results is crucial for reliable wireless networks. Understanding the model's inner workings is key for simplified structures with fewer parameters. To enhance AI-based resource management for trustworthy AI, explainable AI (XAI) has emerged. XAI unveils AI algorithms' inner operations, revealing strengths, weaknesses, and behaviors in future wireless networks. While XAI has been applied in various networking scenarios including DL-based network routing and wireless service provisioning \cite{XAInetworking,khan2024XAImag}, its applicability for radio resource management (RRM) at the physical and MAC layers has been minimally addressed. AI-based RRM relies on complex DL models, which are challenging to comprehend due to their black-box nature. This paper advocates for using XAI techniques to address the black-box nature and uncertainty in predicting the beamforming vectors.

\subsection{Main Contributions}
This work\footnote{A conference version of this work has been accepted to appear in IEEE GLOBECOM 2024 \cite{AsmaaE2E2024Glob}. This version presents the potential of E2E learning for probing and precoding beams but excludes the explainable algorithm. } introduces a pioneering methodology that integrates the joint design of probing beams, and hybrid precoders, forming a comprehensive E2E auto-encoder learning framework tailored for multi-user mmWave systems. {
In the following, we delineate the key contributions of our work}:
{
\begin{itemize}
    \item \textbf{E2E learning of hybrid precoders for multi-user systems:}  A novel methodology for the E2E learning of hybrid precoders tailored for multi-user scenarios is proposed. 
    The initial phase involves the strategic design of probing beams using complex convulational layers tailored to cover specific areas of interest, optimizing their vectors to maximize sensing power in crucial spatial directions. Notably, this optimization adapts the measurement vectors to dynamically changing environments and UE distributions, ensuring ongoing effectiveness. Subsequently, within this learning framework, we utilize the designed probing beams and collected received signal strength indicator (RSSI) information to predict off-grid RF precoders. The overall effective channel, synthesized from the probing beams and UE feedback, serves as the basis for designing digital precoders accounting for multi-user interference.

    \item \textbf{Explainable selection of number of probing beams and bottleneck dimension:} Our method offers an intuitive and explainable approach to selecting the number of probing beams, corresponding to the bottleneck dimension, without treating neural networks as black boxes. In particular, an information-theoretic methodology is introduced aimed at comprehending the learning dynamics and the optimal dimensionality of the bottleneck layer of the proposed autoencoder. This transparency enhances the interpretability and trustworthiness of our model.
    
    \item \textbf{Ease of adoption in cellular standards:} Our proposed methodology is highly adaptable to existing cellular standards, as it does not rely on complex CSI or UE locations. Instead, it solely depends on measuring and reporting a few optimized probing beams, facilitating seamless integration into standard cellular networks. Moreover, the optimized probing beams improve coverage and enable the discovery of new UEs, further enhancing network efficiency.

    \item \textbf{Low beam sweeping latency:} Unlike conventional methods that require sweeping beams for each UE using large predefined codebooks or compressive sensing using a large number of \textit{random} measurement beams, our approach significantly reduces beam sweeping latency. By leveraging site-specific training and generating RF beams in a single-shot, we design a limited set of probing beams, regardless of the number of UEs. With only \textit{8 probing beams}, we outperform the 64-DFT-based codebooks and the 128-oversampled DFT-based codebook, reducing the beam training overhead to  87$\%$ and 93$\%$, respectively.
\end{itemize}
}

\begin{figure}[t]
    \centering
     \includegraphics[width=1\columnwidth]{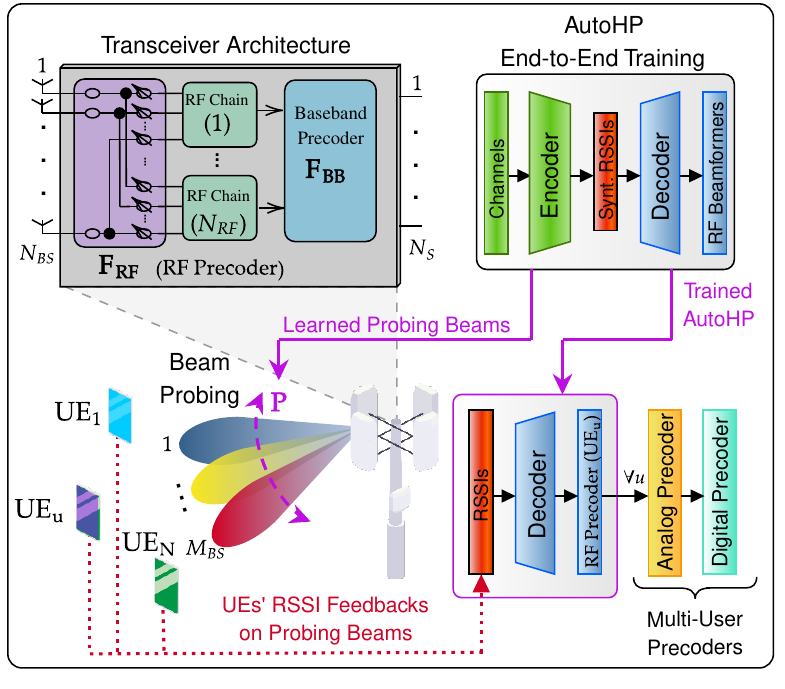}
\caption{System model illustration of the proposed autoencoder-based beam probing and hybrid precoding approach.   }
\label{fig:sys}
\end{figure}

\section{System Model}
\label{s:System_Model}
	{In this section, we detail the adopted system and channel models for hybrid mmWave MIMO system. As shown in Fig.~\ref{fig:sys}, we consider a BS with $\NBS$ antennas and $\NRF$ radio frequency (RF) chains to communicate with $N_{\mathrm{U}}$ single-antenna UEs. We focus on the multi-user beamforming case in which the BS communicates with every UE via only one stream. It is worth mentioning that $N_u = N_{\text{RF}} = N_{\text{streams}}$. This assumption is driven by the spatial multiplexing gain of the described multi-user hybrid precoding system, which is limited by $\min(N_{\text{RF}}, N_u)$ for $N_{\text{BS}} > N_{\text{RF}}$. For simplicity, we will also assume that the BS will use $N_u$ out of the $N_{\text{RF}}$ available RF chains to serve the $N_u$ users. The system is based on a hybrid MIMO architecture of fully-connected network of quantized phase shifters, as described in \cite{Heath2016shiftOrSwitches}.}


\subsection{Channel Model}		
The channel from the BS to the UE$_u$ is expressed based on geometric channel modeling as:
\begin{equation}
    \mathbf{h}_{u} = \sqrt{\frac{\NBS}{L_{\mathrm{}}}} \sum_{\ell=1}^{L_{\mathrm{}}} \alpha_{u,\ell} \mathbf{a}(\theta_{u,\ell},\phi _{u,\ell }),
\end{equation}
where $\alpha_{u,\ell}$ denotes the complex gain of the $\ell$-th path of the $\nth{u}$ UE; $\theta_{u,\ell}$ and $\phi _{u,\ell }$ are the azimuth and elevation angles of departure for the $\ell$-th path, respectively; and $\mathbf{a}(\theta_{u,\ell}, \phi _{u,\ell })$ represents the steering vector. 
For an $ N_{h}\times N_{v}$ uniform planar array (UPA) with a total of $N_{\mathrm{t}} =N_{h} N_{v}$ antenna elements, ${\mathbf{a}}\left ({\vartheta,\psi }\right)$ can be represented by 
\begin{align}
	{\mathbf{a}}\left ({\theta,\phi }\right) \!=\! \frac {1}{\sqrt {\NBS} }{\left [{\! {{e^{ - j2{\pi }d_1(\theta,\psi) {\mathbf{n}}_{1}/{\lambda }}}} \!}\right]}\!{\otimes }\!{\left [{ {{e^{ - j2{\pi }d_2(\phi) {\mathbf{n}}_{2}/{\lambda }}}} }\right]}\!, 
\end{align}
where ${\mathbf{n}}_{1}=[0,1,\cdots,N_{h}-1]$, ${\mathbf{n}}_{2}=[0,1,\cdots,N_{v}-1]$, $\lambda$ is the carrier wavelength, $d_1(\theta,\phi)=d\sin(\theta)\cos(\phi)$, $d_2(\phi)=d\sin(\phi)$, and $d$ is the antenna spacing typically satisfying $d = \lambda /2$. Notably, for mmWave frequencies, channels often exhibit sparsity in the angular domain, leading to a constrained number of channel paths, $L$. 

\subsection{Initial Access}

During the beam alignment/probing process, as depicted in Fig. \ref{fig:beamConv}, the BS scans a limited number of probing beams to gather information about the channels $\mbf{h}_u$ for $N_U$ UEs. Specifically for beam alignment, processing in the RF domain is only considered, where  it is assumed that the BS utilizes a single RF chain connected to the phase shifters of all antenna elements.\footnote{{While multiple RF chains can enhance SNR, a single RF chain is often sufficient for establishing the initial connection. Our approach can leverage multiple RF chains at the receiver to reduce beam training overhead by directly applying the designed probing beams, with each RF chain implementing one of the beams. Moreover, during subsequent data communication stages, multiple RF chains are used to enable multi-user simultaneous transmission, improving system capacity and performance.  }} Let $\mbf{P}\in \mathbb{C}^{\NBS \times \MBS}$ be the probing matrix adopted by the BS, with entries adhering to the constant modulus constraint for analog beamforming, where $\MBS$ is defined as the number of probing beams inside $\mbf{P}$. The BS periodically transmits probing symbols $s_p$ through the beams defined by the matrix $\mbf{P}$. This cyclic sweeping of probing beams during SSBs facilitates the discovery and synchronization of new UEs with the BS. 


Following the beam sweeping process,  the composite received signal at the UE$_u$ $\mbf{r}_{\mathrm{RSSI},u} \in \mathbb{C}^{1\times \MBS}$ comprising the received signals from all probing beams, can be expressed as
\begin{equation}\label{equ:recRSSIu}
       \mbf{r}_{\mathrm{RSSI},u} = \sqrt{P_t} {\mbf{h}}_{\,u}^{\mathsf{H}}{\mbf{P} s_{p}} + {\mbf{n}}_{\mathrm{RSSI},u}, 
\end{equation}
where ${\mbf{n}}_{\mathrm{RSSI},u} \in \mathbb{C}^{1\times \MBS}$ is the additive white Gaussian Noise (AWGN) measurement noise vector whose each element has noise power $\sigma^2_{p}$.
 We model the effective SNR of the $\nth{m}, \forall m \in [1,\MBS]$ probing measurements for the $\nth{u}$ UE as follows:
\begin{equation}
    \eta_{\mathrm{RSSI},u, m}=  \frac{P_t |\mbf{h}_u^{\mathsf{H}} \mbf{p}_m|^2}{\sigma^2_p}.
\end{equation}
The BS sweeps $\MBS$ probing beams, then the UE reports the $\MBS$ received signal power measurements defined as
\begin{equation}\label{equ:RSSI}
    \mbf{y}_{\mathrm{RSSI},u}=\left[|[\mbf{r}_{\mathrm{RSSI},u}]_1|^2 \cdots |[\mbf{r}_{\mathrm{RSSI},u}]_{\MBS}|^2 \right].
\end{equation}
Notably, the overall number of beams swept remains constant, unaffected by the number of UEs, as multiple UEs can measure the probing beams concurrently, utilizing shared time and frequency resources. It is worth noting that conventional probing beams (e.g., DFT-based codebooks) may have low SNR values due to their {quantized} coverage in angular space. To address the low SNR in the probing measurement phase, we leverage the exploration capabilities of DL 
to identify optimal probing beams tailored to specific sites. 
\subsection{Data Transmission}

During data transmission in the downlink, the BS adopts an  RF beamformer $\FRF= [\mathbf{f}^{\mathrm{RF}}_1, \mathbf{f}^{\mathrm{RF}}_2, \dots, \mathbf{f}^{\mathrm{RF}}_{\NRF}] \in \mathbb{C}^{\NBS \times \NRF}$ and  a baseband (BB) precoder $\FBB =[\mathbf{f}^{\mathrm{BB}}_1, \mathbf{f}^{\mathrm{BB}}_2, \dots, \mathbf{f}^{\mathrm{BB}}_{N_U}] \in \mathbb{C}^{\NRF \times N_{U}}$, the transmitted signal is then given by
\begin{equation}
    \mathbf{x}=\FRF \FBB \mathbf{s}
\end{equation}
where $\mathbf{s}=[s_1,s_2, \dots,s_{N_U}]^\mathrm{T}$  is an $N_U \times 1$ vector representing the symbols transmitted to $N_U$ UEs, satisfying the condition $\expec{\mathbf{s}\mathbf{s}^\mathrm{H}}= \tfrac{P}{N_U}\mathbf{I}_{N_U}$, where $P$ denotes the average total power.  Furthermore, we consider that the angles of the analog phase shifters are quantized and belong to a finite set of possible values, yielding $[\FRF]_{m,}=\tfrac{1}{\sqrt{\NBS}}\exp{j\theta_{m,n}}$, where $\theta_{m,n}$ are quantized phases.  The total power constraint is maintained by normalizing  $\FBB $ where $\|\FRF \FBB\|^2_F=N_U$. Consequently, the UE$_u$ receives the signal as
\begin{equation}\label{equ:rec_u}
       {{r}}_{u}={\mbf{h}}_{\,u}^{\mathsf{H}}\sum_{n=1}^{N_U}{\FRF {\mbf{f}}^\mathrm{\,BB}_{\,n}s_{n}} + {{n}}_{\,u}, 
\end{equation}
where ${{n}}_{\,u} \sim \mathcal{N} ({0}, \sigma^2 )$ is the Gaussian noise corrupting the received signal.  Therefore, the signal-to-interference plus-noise ratio (SINR) achieved is given by
\begin{equation}
    \eta_u= \frac{\tfrac{P}{N_U} |\mbf{h}_u^{\mathsf{H}} \FRF \mbf{f}^{\mathrm{BB}}_u|^2}{\tfrac{P}{N_U} \sum_{n\neq u}|\mbf{h}_u^{\mathsf{H}} \FRF \mbf{f}^{\mathrm{BB}}_n|^2 + \sigma^2}.
\end{equation}
 
\section{Problem Formulation}

Our primary goal is to optimize the design of the analog (i.e., RF) and digital (i.e., baseband) precoders at the BS in order to maximize the overall sum-rate of the system. Following from the received signal at the UE$_u$ in \eqref{equ:rec_u}, the achievable rate of the UE$_u$ is expressed as
\begin{equation}\label{eq:rate}
R_u=\log_2\left(1+\frac{\tfrac{P}{N_U} |\mbf{h}_u^{\mathsf{H}} \FRF \mbf{f}^{\mathrm{BB}}_u|^2}{\tfrac{P}{N_U} \sum_{n\neq u}|\mbf{h}_u^{\mathsf{H}} \FRF \mbf{f}^{\mathrm{BB}}_n|^2 + \sigma^2} \right),
\end{equation}
resulting a network sum-rate of $R_{\mathrm{sum}}=\sum_{\forall u} R_u$.
{Due to the limitations of RF hardware, the the phase shifts of  the analog beamforming vectors are limited to a subset of $2^b$ discrete values uniformly distributed in the range $(-\pi, \pi]$, where $b$ denotes the quantization level. Consequently, these vectors must be selected from a massive discrete quantized search space denoted by $\mathcal{F}$. To illustrate, for a BS equipped with $32$ elements and employing $3$-bit phase quantization, the search space {$\mathcal{F} $} comprises of $7.9\times 10^{28}$ potential beamforming vectors.} 

If the system sum-rate is chosen as a performance metric, the precoding design problem entails finding solutions for $\FRF$, and $\{\mbf{f}^{\mathrm{BB}}_u\}_{u=1}^{N_U}$ that solve
\begin{align}\label{equ:probA}
&\left\{{\FRF}^{\,\star},\left\{{\mbf{f}}_{\,u}^{\,\star\textrm{BB}}\right\}_{u=1}^{N_{U}}\right\}\nonumber\\ 
    &\ \ \ =\arg\max{\sum_{u=1}^{N_{U}}\log_2\left(1+\frac{\tfrac{P}{N_U} |\mbf{h}_u^{\mathsf{H}} \FRF \mbf{f}^{\mathrm{BB}}_u|^2}{\tfrac{P}{N_U} \sum_{n\neq u}|\mbf{h}_u^{\mathsf{H}} \FRF \mbf{f}^{\mathrm{BB}}_n|^2 + \sigma^2}\right)}\nonumber\\ 
    &\qquad \mbox{s.t.}\qquad\left[{\FRF}\right]_{:,u} \in\mathcal{F} , u=1,2,\ldots, N_{U},\nonumber\\ 
    &\qquad\qquad\quad\left\|{\FRF}\left[{\mbf{f}}_{\,1}^\textrm{\,BB}, {\mbf{f}}_{\,2}^\textrm{\,BB},\ldots,{\mbf{f}}_{\,N_{U}}^\textrm{\,BB} \right]\right\|_F^2=N_{U},
\end{align}
which is a mixed-integer programming problem whose solution involves an exhaustive search across the entire space $\mathcal{F} $ encompassing all potential combinations of $\FRF$. Additionally, the joint design of the digital precoder $\FBB$ is necessary, entailing the design of analog beamforming/combining vectors. In practice, this may involve providing feedback on the channel matrices $\mbf{h}_u, \forall u \in [1,N_U]$, or the effective channels, $\mbf{h}_u^{\mathsf{H}} \FRF$. Consequently, solving \eqref{equ:probA} imposes a substantial training and feedback overhead. Moreover, determining the optimal digital linear precoder is generally unknown, even without RF constraints, and only iterative solutions exist \cite{Poor2020MUBF, khateeb2015MUBF}. As a result, directly solving this sum-rate maximization problem is neither practical nor feasible. 

In classical (non machine learning) signal processing, there are primarily two solutions in designing the beamforming vectors which are defined in the Section \ref{sec:Intro}: 1) \textbf{Channel state information  based beamforming techniques} and 2) \textbf{Beam Sweeping-based techniques}. The main limitation of these approaches is that they do not leverage from prior channel observations for minimizing the training overhead related to channel estimation and precoder design. 
To address this challenge, the aim of our paper is to create a unified learning framework. This framework directly learns a small-sized, site-specific probing beam codebook and identifies the user-specific RF precoding vectors for the hybrid architecture. Consequently, it minimizes channel training overhead without requiring extensive searches through large codebooks.
In the subsequent section, we illustrate how tools from machine/deep learning can offer an effective resolution to this problem.

\begin{figure}
    \centering
    \includegraphics[width=0.99\linewidth]{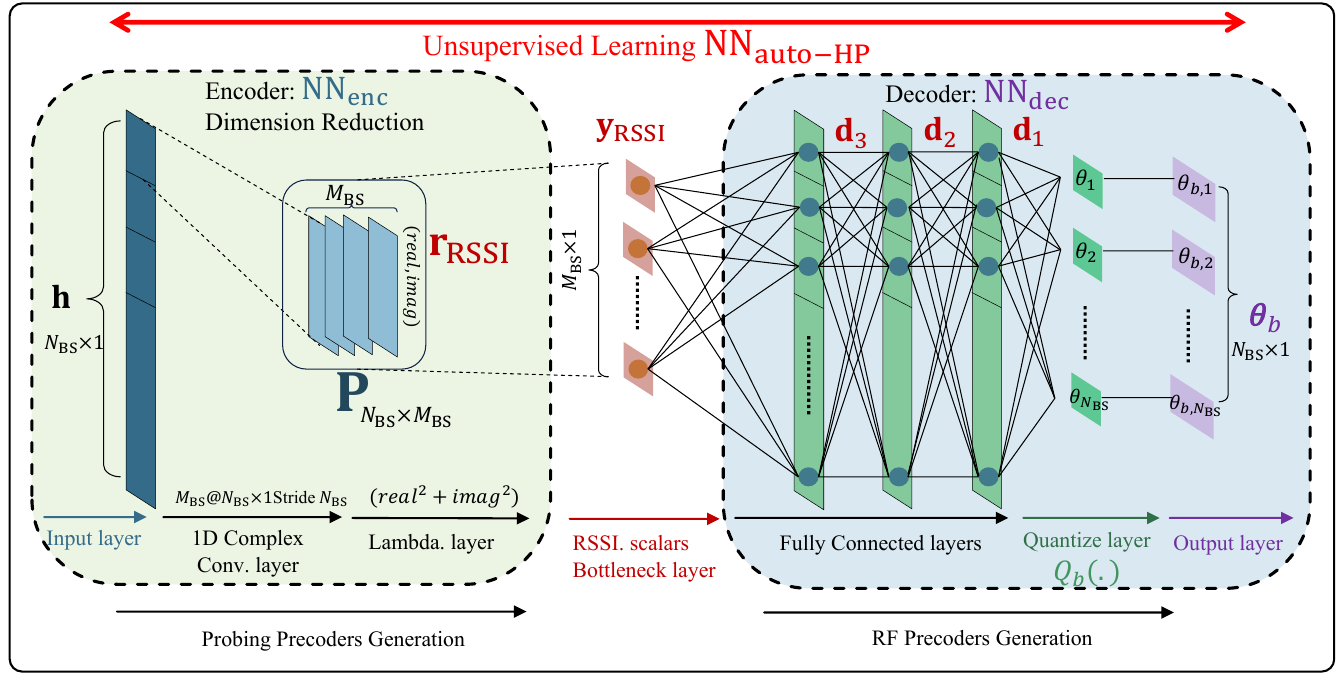}
    \caption{The proposed autoencoder NN$_{\text{Auto-HP}}$ consists of two sections: the probing precoders generator (encoder NN$_{\text{enc}}$) and the RF precoders generator (decoder NN$_{\text{dec}}$), where $  \mbf{r}_{\mathrm{RSSI}}$, $\mbf{y}_{\mathrm{RSSI}}$, $\mbf{d}_{3}$, $\mbf{d}_{2}$, $\mbf{d}_{1}$ are the outputs of each hidden layer in NN$_{\text{Auto-HP}}$.}
    \label{fig:NetArch}
\end{figure}

\section{AutoHP-Based Beam Probing and Multi-User Hybrid Precoding}

Addressing \eqref{equ:probA} presents an extra challenge due to the division of the precoding operation into distinct analog and digital domains, 
each characterized by different constraints. 

The fundamental concept behind the proposed algorithm is to decompose the computation of the precoders into multiple stages, as illustrated in Fig.\ref{fig:sys}. First, an E2E learning framework is introduced to jointly learn the initial access probing beams and to train an autoencoder that predicts the optimal 
RF precoders, which are designed to maximize the desired signal power of each UE, neglecting the resulting interference among UEs. 
Within this E2E learning framework, a concise set of probing beams ($\mbf{P}$) is learned at the BS. The received signal power measurements of these probing beams are then utilized as inputs to its decoder,  specifically  RF precoding beam generator $g(.)$, to synthesize an optimal narrow RF beam from the \textbf{continuous} search space. 
{Then, we introduce a customized loss function that maximizes the desired signal power at UEs, as defined in \eqref{equ:probB}, in an unsupervised learning manner while also maximizing the mutual information (MI) at the compressed bottleneck layer.} Finally, once the training is complete, we develop a multi-user hybrid precoding technique that utilizes the trained autoencoder 
to predict the RF precoders for all UEs and design the digital BB precoders while mitigating multi-user interference. 

Our method attains multi-fold gains: (i) Unlike the common use of generic and non-optimized beam codebooks like DFT-based codebooks, or even random probing vectors commonly employed in traditional channel compressing sensing methods, our approach focuses on learning the optimization of probing vectors. The autoencoder adapts to the UE distribution and environmental factors, enabling a targeted concentration of measurement/sensing power on the most promising spatial directions. (ii) The deep learning model ultimately learns to directly predict hybrid beamforming vectors from the compressed probing measurements.

\subsection{Offline Training: E2E Learning of Probing and Precoding}

The proposed E2E learning framework designs the RF beamformers to maximize the desired signal power of a particular UE while neglecting  the other UEs' interference.   The BS finds ${\mbf{f}^{\mathrm{RF}}_u}^{\star}$ for each UE$_u, \forall u,$ by solving
 \begin{align}\label{equ:probB}
 \displaystyle {\mbf{f}^{\mathrm{RF}}_u}^\star &=  \max_{\forall{\mbf{f}^{\mathrm{RF}}_u}  \in \mathcal{F}}{|\mbf{h}_u^{\mathsf{H}} {\mbf{f}^{\mathrm{RF}}_u}|^2}
 \\
     &\textrm {s.t.  } {\mbf{f}^{\mathrm{RF}}_u}  = g(\mathbf {y}_{\mathrm{RSSI},u}),\notag
     \\
     &\hphantom {\textrm {s.t.} }|\mathbf {[P]}_{i,j}| = \frac {1}{\sqrt {\NBS}}, \forall i\in [1,\NBS], \forall j=[1,\MBS] , \notag 
 \end{align}
which is non-convex and difficult to solve due to the unit-modulus constraints of the probing beamforming weights of $\mbf{P}$ and the unknown generator function $g(.)$. We propose optimizing both the probing beams $\mbf{P}$ and the RF precoding beam generator using E2E learning framework as explained in the sequel.\footnote{In this section, we will omit the subscript $u$ since the offline training procedure is tailored for a specific UE and repeated for all UEs.} {It is worth noting that given the enormous search space, treating the problem as a classification task would involve categorizing each possible phase shift into one of the possible classes, which is computationally infeasible and inefficient. Instead, by formulating the problem as a regression task, we can predict the continuous values of the phase shifts, which can then be quantized to the nearest discrete value.}

\subsubsection{\textbf{Neural Network Architecture}}
We introduce our innovative autoencoder network architecture, denoted as NN$_{\text{Auto-HP}}$, depicted in Figure \ref{fig:NetArch}. NN$_{\text{Auto-HP}}$ is tasked with two main objectives: (1) generating the probing beams matrix $\mbf{P}$ through the encoder part of the NN, NN$_{\text{enc}}$, and (2) generating the RF phase shifts through the decoder part of the NN, NN$_{\text{dec}}$. Examining the received sensing signal formulation in \eqref{equ:recRSSIu}, it is observed that the emulation of the product of the channel vector $\mbf{h}$ and the probing matrix $\mbf{P}$ can be achieved by inputting this channel vector into NN$_{\text{enc}}$, which comprises a complex convolutional layer and a custom lambda layer. It is important to note here that since the transmitter probing weights are complex, we adopt the complex-valued neural network implementation of the convolutional layers in \cite{Trabelsi2017DeepCN}, we provide more details on how the probing matrix is constructed in the Section \ref{sec:ProbD}. The NN$_{\text{enc}}$ module essentially computes complex-valued matrix multiplication and addition on the channel vector. The output of the complex convolutional layer has $\MBS$ feature maps. To enforce the unit-modulus constraint, the complex matrices are normalized element-wise by the magnitude.  The complex convolutional layer computes the composite vector $  \mbf{r}_{\mathrm{RSSI}}$ of received signals of all  probing  beams in \eqref{equ:recRSSIu} and can be considered to perform virtual sweeping of the probing beam pairs while being differentiable. Then, the complex vector $  \mbf{r}_{\mathrm{RSSI}}$ is fed into the lambda layer that calculates its amplitude and produce $\mbf{y}_{\mathrm{RSSI}}$ in  \eqref{equ:RSSI}.

The power measurements corresponding to the $\MBS$ probing beams, denoted as $\mbf{P}$, along with their elements represented by $\mbf{y}_{\mathrm{RSSI}}$, serve as inputs for the RF precoding beam generator NN$_\text{dec}$. This NN architecture is configured with a multilayer perceptron (MLP) regression, featuring three hidden layers and a final linear layer that produces the RF synthesized phase shifts. Subsequent to each fully connected layer, a sequence of operations including rectified linear unit (ReLU) activation, batch normalization, and a dropout layer is applied. The outputs from these hidden layers are labeled as $\mbf{d}_1$, $\mbf{d}_2$, and $\mbf{d}_3$, respectively. 
The hidden layers and the final layer of the MLP have dimensions equivalent to the number of BS transmit antennas, $\NBS$. As a result, the final layer of the MLP generates predicted phase shifts prior to quantization. Subsequently, these predicted phase shifts are directed to a quantization layer denoted as ${Q}_b(.)$, which quantizes the phase shifts to valid phases using $b$-bits quantization.

The loss is computed by evaluating the beamforming gain of the synthesized beams and the probing beamforming vector for each batch of training channel realizations, as elaborated in Section \ref{sec:loss}. Both the NN$_\text{enc}$ and NN$_\text{dec}$ are trained using stochastic gradient descent and backpropagation algorithms. Prior to deployment, the NN$_{\text{Auto-HP}}$ model undergoes offline training, utilizing training data obtained from ray-tracing simulations or measurements. 

\subsubsection{\textbf{Probing Beam Codebook Design}}\label{sec:ProbD}

The process of beam sweeping utilizing the probing codebook is represented by a complex module within the NN$_{\text{enc}}$ architecture, which calculates the complex received beamforming signals along with their corresponding power. The input to this NN module is the BS-UE's channel vector $\mbf{h}$ ($\mbf{h}_u$). The complex layer in NN$_{\text{enc}}$ executes the complex arithmetic associated with analog beamforming using real arithmetic. In the parameterization of a $\MBS$-beam codebook $\mbf{P}$, the trainable weights of the complex layer constitute elements of $\boldsymbol{\Phi} \in \mathbb{R}^{\NBS \times \MBS}$, representing the phase shift values applied to each antenna element. The computation of the complex beamforming weights $\mathbf{P}=[\mathbf{p}_1,\cdots,\mathbf{p}_{\MBS}] \in \mathbb{C}^{\NBS \times \MBS}$ can then be expressed as
\begin{equation}\label{equ:prob}
\mbf{P} = \frac {1}{\sqrt {\NBS}}(\cos \boldsymbol{\Phi}+j\cdot \sin \boldsymbol{\Phi}).
\end{equation}
The matrix multiplication equivalent to the complex matrix multiplication ($ \mbf{r}_{\mathrm{RSSI}} = \mathbf{P}^{\mathsf{H}} \mathbf{h}$) can be expressed in terms of real and imaginary components as:
\begin{align} 
    \begin{bmatrix} 
        \mathbf{r}^{real}_{\mathrm{RSSI}} \\ \mathbf{r}^{imag}_{\mathrm{RSSI}} \end{bmatrix} = \begin{bmatrix} \mbf{P}^{real} &\quad -\mbf{P}^{imag}\\ \mbf {P}^{imag} &\quad \mbf{P}^{real} \end{bmatrix}^{\mathrm{T}} \begin{bmatrix} \mathbf {h}^{real} \\ \mathbf {h}^{imag}
    \end{bmatrix},
\end{align}
where $\mathbf{h} \in \mathbb{C}^{\NBS \times 1}$ is the channel vector of a particular UE, $\mathbf{r}_{\mathrm{RSSI}} \in \mathbb{C}^{\MBS \times 1}$ is the beamforming output. Then, beamforming signal power expressed as
\begin{align} 
\mbf{y}_{\mathrm{RSSI}}= \left [{( {r}_{1}^{real})^{2}+( {r}_{1}^{imag})^{2},\cdots,( {r}_{\MBS}^{real})^{2}+( {r}_{\MBS}^{imag})^{2}}\right]^{T}.  
\end{align}
Although $[\mbf{y}_{\mathrm{RSSI}}]_{j}$ is not complex differentiable with respect to $r_{j}$, its gradient can be computed as $\frac {\partial |r_{j}|^{2}}{\partial r_{j}} =\begin{bmatrix} \frac {\partial |r_{j}|^{2}}{\partial r^{real}_{j}},\frac {\partial |r_{j}|^{2}}{\partial r^{imag}_{j}} \end{bmatrix}$  using backpropagation by treating the real and imaginary components of $r_{j}$ independently. The NN$_{\text{enc}}$ weights $\boldsymbol{\Phi}$ can then be adjusted using the chain rule and backpropagation.

 Given that during the NN initialization, only the dimension of $\boldsymbol{\Phi}$ needs specification (i.e. $\MBS$ and $\NBS$), NN$_{\text{enc}}$ merely requires knowledge of the number of antenna elements and the number of probing beams, not the array geometry. The array-geometry details are integrated into the input channel vectors, enabling NN$_{\text{enc}}$ to autonomously acquire the optimal probing beam phases for each antenna element. This adaptability ensures that the architecture can be easily adopted by BSs featuring diverse antenna arrays.

It's important to highlight that through E2E training, NN$_{\text{enc}}$ undergoes unsupervised learning to optimize its probing beam matrix (or kernel weights). Hence, this optimization tailors the the probing beam matrix $\mbf{P}$ to the surrounding environment and UE distribution, concentrating the sensing power on promising spatial directions.

\subsubsection{\textbf{RF Precoding Beam Phase Shifts Design}}
Utilizing the received RSSI vector $\mbf{y}_{\mathrm{RSSI}}$, a DNN is employed to learn the direct mapping function $g(.)$ from $\mbf{y}_{\mathrm{RSSI}}$ to the RF beamforming vectors $\mbf{f}^{\mathrm{RF}}$. It's worth emphasizing that the significant training overhead in mmWave systems primarily arises from determining these RF beamforming vectors. Once the RF beamforming vectors are determined, the low-dimensional ($1 \times \MBS$) effective channel $\mbf{h}_u^{\mathsf{H}}\FRF$ becomes easily estimable, simplifying the construction of BB precoders. Given that the RF beamforming vectors are chosen from a quantized legitimate set $\mathcal{F}$, we formulate the problem of predicting the phase shifts of these RF beamforming vectors as a regression problem.

The beam generation function $g(.)$, represented by NN$_{\text{dec}}$, takes the power of complex received signals of all beams in $\mbf{P}$ as input ($\mbf{y}_{\mathrm{RSSI}}$), computed using NN$_{\text{enc}}$ during training or through beam sweeping during deployment.  An evident strategy for optimizing $g(.)$ is to design it to predict the optimal narrow beam ${\mbf{f}^{\mathrm{RF}}}^\star$, maximizing SNR with the current channel, as adopted in the loss function in Section \ref{sec:loss}. 
The RF beamformer is modeled using the phase outputs as follows
\begin{align}
    {\mbf{f}^{\mathrm{RF}}}=& \frac{1}{\sqrt{\NBS}} e^{j{\boldsymbol\theta}_b} 
\end{align}
where ${\boldsymbol\theta}_b= Q_b({\boldsymbol\theta}) $.
The MLP, being a potent function approximator, can yield accurate estimates of quantized phases. By designing the narrow beam with the highest SNR for the $\nth{u}$ UE, the BS can enhance beam alignment robustness, directly utilizing the MLP output and reducing the search space by avoiding extensive beam training overhead. 

\subsection{Loss Function Design}\label{sec:loss}


In situations where a significant number of parameters require optimization, leveraging unsupervised learning becomes advantageous due to its inherent flexibility. This involves the formulation of a custom loss function \cite{nikbakht2021unsupervised-parametric-optimization} tailored to address the optimization problem.  In this section, we tailor a loss function for the problem defined in \eqref{equ:probB} while concurrently aiming to maximize the entropy at the bottleneck layer, as will be expressed in \eqref{equ:explainEntropy} in Section \ref{sec:entropyprof}.

 The pursuit of maximizing the signal power objective in \eqref{equ:probB} and entropy of the bottleneck layer  can be formulated as the minimization of the loss function  that does not require explicit labels for the probing beams is defined as
 \begin{equation}
     \mathcal{L}_{\text{max-pwr-entropy}}= - \frac{1}{|{\mathcal{H}_U^{B}}|}\sum_{{\mbf{h}}_u \in \mathcal{H}_U^b} (    {|\mbf{h}_u^{\mathsf{H}} {\mbf{f}^{\mathrm{RF}}_u}|^2}+ H( \mbf{y}_{\mathrm{RSSI}})),
 \end{equation}
 where ${\mbf{f}^{\mathrm{RF}}_u}= \frac{1}{\sqrt{\NBS}} e^{j{\boldsymbol\theta}_{b,u}} $,  ${\mathcal{H}_U^{B}}$ is the set of training channel batch samples such that $B$ is the batch size.  $H( \mbf{y}_{\mathrm{RSSI}})$ is the entropy of the bottleneck layer which will be defined in Section \ref{sec:entropyprof} in  \eqref{equ:explainEntropy} as we aim to emphasize the information of the bottleneck by maximizing its entropy.
	\begin{algorithm}[t!]
		\caption{ \textsc{\small{E2E learning Multi-User Hybrid Precoders}}}\label{Alg:1}
  \footnotesize
			\begin{algorithmic}[1]
            \State{\textbf{\color{gray} Offline learning Mode:}}
            \State {\textbf{{Initialize}}: Channel samples: $\mathcal{H}_U =  [\mbf{h}_i]_{i=1}^{N_D} $ where $N_D$ is the number of channel samples. \label{Alg1:line:1}}

                     $\mbf{P}$, NN$_{\text{dec}} \leftarrow${\small{\textsc{E2E Learning Probing and RF precoding}}}{$(\mbf{h}_i \in \mathcal{H}_U \forall i )$}
                     \vspace{2pt}
				\hrule 
				\vspace{2pt}
                     \State{\color{teal} \textbf{Online Deployment Mode:}}
                     
				 {\!\!\!\!\!\!\!\!\!\!\!\textbf{\color{teal}S1}: Beam Sweeping over the probing beams}
				\State $\mbf{y}_{\mathrm{RSSI},u}\leftarrow$ Beam Sweep over $\mbf{P}$ and collect RSSIs $\forall u$
    
                {\!\!\!\!\!\!\!\!\!\!\!\textbf{\color{teal}S2}: Feed back $\mbf{y}_{\mathrm{RSSI},u}, \forall u$ to BS}
                
                {\!\!\!\!\!\!\!\!\!\!\!\textbf{\color{teal}S3}: Generate RF precoders per UE $u$}
                      \State ${\mbf{f}^{\mathrm{RF}}_{u}}^{\star}\xleftarrow[{\mathrm{ NN}_{\text{dec}}}]{\text{Online}} \mbf{h}_u, \quad \forall u$  
                      \State BS sets $\FRF\leftarrow[{\mbf{f}^{\mathrm{RF}}_{1}}^{\star}, {\mbf{f}^{\mathrm{RF}}_{2}}^{\star}, \dots, {\mbf{f}^{\mathrm{RF}}_{N_U}}^{\star}]$
                      
			 {\!\!\!\!\!\!\!\!\!\!\!\textbf{\color{teal}S4}: Generate multi-user BB precoders}
				\State {$\FBB \leftarrow {\small{\textsc{multi-user BB precoding}}}({\FRF, \mbf{h}_u, \forall u})$}
    
    {\!\!\!\!\!\!\!\!\!\!\!}\Return $\FRF,\FBB$
                     \vspace{2pt}
				\hrule 
				\vspace{2pt}
				
				\Procedure{\small{E2E Learning Probing and RF precoding}}{$\mbf{h}_i \in \mathcal{H}_U \forall i\in [1,N_D] $}
				\State ${\mbf{f}_{i}^{\mathrm{RF}}}^{\star}\xleftarrow[{\mathrm{ NN}_{\text{Auto-HP}}}]{\text{Offline train}} \mbf{h}_i, \quad \forall i$  
    
    where  ${\mathrm{ NN}_{\text{Auto-HP}}}=[\mathrm{ NN}_{\text{enc}}\mathrm{ NN}_{\text{dec}}]$
                    \State $\mbf{P}\leftarrow $ Extract Weights $(1^{\mathrm{st}}$ layer NN$_{\text{enc}})$ // 
                    as per \eqref{equ:prob}
				\Return $\mbf{P}$, NN$_{\text{dec}}$
				\EndProcedure
				\vspace{2pt}
				\hrule 
				\vspace{2pt}
				
				\Procedure{\small{multi-user BB precoding}}{$\FRF, \mbf{h}_u, \forall u$}
				\For {$u=1:N_U$}
                        \State UE estimates its effective channel
                        \State ${{\widetilde{\mbf{h}}_u}}=\mbf{h}_u^{\mathsf{H}}\FRF$
                        \State   $\widehat{{\mbf{h}}}_{u}\leftarrow\mathop{\arg\max}\limits_{\hat{{\mbf{h}}}_u \in \mathcal{H}}{\|{{\tilde{\mbf{h}}_u}}^{\mathsf{H}} \widehat{{\mbf{h}}}_{u}\|}$ // UE $u$ quantizes ${{\widetilde{\mbf{h}}_u}}$ 
                        
                        \State UE feeds back $\widehat{\mbf{h}}_u$ 
                        
                    \EndFor\\
                    \State  $\FBB\leftarrow {\widehat{{\mbf{H}}}}^{\mathsf{H}}\left(\widehat{{\mbf{H}}} {\widehat{{\mbf{H}}}}^{\mathsf{H}}\right)^{-1}$, // BS designs $\FBB$,  where $$\widehat{{\mbf{H}}}=\left[\widehat{{\mbf{h}}}_{\,1},\ldots,\widehat{{\mbf{h}}}_{\,u}, \ldots, \widehat{{\mbf{h}}}_{\,N_U}\right]^{\mathsf{H}}$$
                    \State BS normalizes ${\mbf{f}}_{\,u}^\textrm{\,BB}\leftarrow\frac{{\mbf{f}}_{\,u}^\textrm{\,BB}}{\|{\mbf{F}}_\textrm{RF} {\mbf{f}}_{\,u}^\textrm{\,BB}\|_F}, \forall u\in [1,N_U]$\\
				\Return $\FBB$
				\EndProcedure
			\end{algorithmic}
	\end{algorithm}
 
\subsection{Online Deployment: multi-user Hybrid Precoding Design}

{ Once training is complete, we switch to deployment mode.  During deployment, we \textbf{no longer rely on prior channel measurements} to predict the RF precoders. Instead, the trained weights of the NN$_{\text{enc}}$ $\mathbf{\Phi}$ are first utlized to construct the probing codebook matrix $\mathbf{P}$ following \eqref{equ:prob}. The probing vectors from $\mathbf{P}$ are used to probe the environment and collect RSSIs from UEs, where these RSSIs will be used by the trained NN$_{\text{dec}}$ to predict the RF phase shifts for each UE. Finally, the BB precoders are designed for the multi-UEs while considered the inter-user interference. In the following, we delineate the online deployment process in four sequential stages:}
\begin{enumerate}
    \item[\textbf{S1}] Once the model is trained, the kernels of NN$_{\text{enc}}$ is directly employed as the probing beam matrix $\mbf{P}$ by the BS during the beam sweeping stage.
    \item[\textbf{S2}] The RSSI measurements, $\mbf{y}_{\mathrm{RSSI},u}, \forall u$, reported by the UEs are collected. 
    \item [\textbf{S3}] The reported sensing vector $\mbf{y}_{\mathrm{RSSI},u}$ is fed into the trained RF precoder network NN$_{\text{dec}}$ in Fig. \ref{fig:NetArch}, represented in the second segment of NN$_{\text{Auto-HP}}$, to predict the RF beamforming vectors for each UE and finally forming the RF beamforming matrix $\FRF$.
    \item[\textbf{S4}] Once the optimal RF beamformers for each UE are obtained, i.e., $\FRF=[{\mbf{f}^{\mathrm{RF}}_{1}}^{\star}, {\mbf{f}^{\mathrm{RF}}_{2}}^{\star}, \dots, {\mbf{f}^{\mathrm{RF}}_{N_U}}^{\star}]$, the BS utilizes the effective channels, denoted as $\tilde{\mbf{h}}_u = {\mbf{h}}_u\FRF, \forall u$, in collaboration with UEs to formulate the baseband precoders. Notably, the dimension of each effective channel vector $\tilde{\mbf{h}}_u$ is $\NRF \times 1$, \textbf{significantly smaller than the original $\NBS \times 1$ channel vector \cite{khateeb2015MUBF}.} Subsequently, each UE quantizes its effective channel by utilizing a designated codebook $\mathcal{H}_B$ and transmits the index of the quantized channel vector back to the BS using $B_{H}$ bits, where quantized channel for the UE$_u$ is represented as $\widehat{\mbf{h}}_u $. Ultimately, the BS devises its zero-forcing (ZF) digital precoder based on the received quantized channels. The adoption of narrow beamforming and the sparse nature of mmWave channels is anticipated to result in a well-conditioned effective MIMO channel, enabling the utilization of a straightforward multi-user digital beamforming strategy such as ZF that can achieve performance close to optimality \cite{khateeb2015MUBF}. Therefore, the designed BB precoder is expressed as 
\begin{equation}
   \FBB= {\widehat{{\mbf{H}}}}^{\mathsf{H}}\left(\widehat{{\mbf{H}}} {\widehat{{\mbf{H}}}}^{\mathsf{H}}\right)^{-1},
\end{equation}
where $\widehat{{\mbf{H}}}=\left[\widehat{{\mbf{h}}}_{\,1},\ldots, \widehat{{\mbf{h}}}_{\,u}, \ldots,\widehat{{\mbf{h}}}_{\,N_U}\right]^{\mathsf{H}}$. Then, the BS  normalizes the BB precoders for all UEs ($\forall u\in [1,N_U]$) as follows
\begin{equation}
    {\mbf{f}}_{\,u}^\mathrm{\,BB}=\frac{{\mbf{f}}_{\,u}^\mathrm{\,BB}}{\|{\mbf{F}}_\textrm{RF} {\mbf{f}}_{\,u}^\mathrm{\,BB}\|_F}.
\end{equation}
\end{enumerate}

\begin{figure*}
    \centering
    \includegraphics[width=0.7\linewidth]{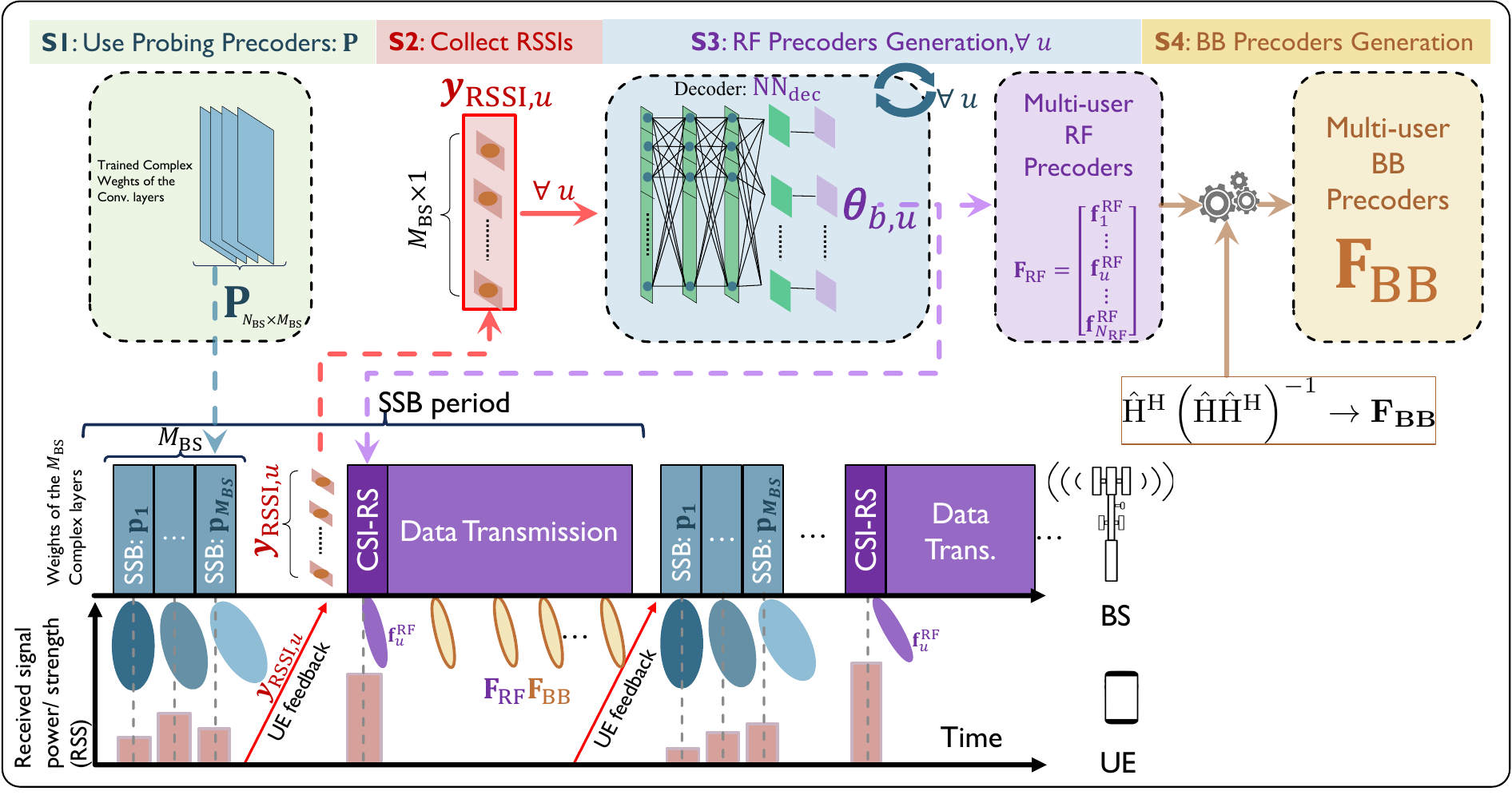}
    \caption{Illustration of the Deployment mode.}
    \label{fig:DepMode}
\end{figure*}

Prior works  \cite{li2019deep,Andrews2022BF,heng2023grid} delved into scenarios of single-UE single-stream MIMO systems, where analog precoders were devised to optimize either the received signal strength or the sum-rate. In contrast, our study presents a distinct setup involving multi-user downlink transmission. Consequently, the primary objective of our hybrid analog/digital beamforming approach differs from that in \cite{li2019deep,Andrews2022BF,heng2023grid}, as we specifically address the challenge of managing multi-user interference in our proposed framework. The algorithm of the proposed multi-stage E2E learning-based multi-user hybrid precoding scheme is summarized in Algorithm \ref{Alg:1} and illustrated in Fig. \ref{fig:DepMode}.

\section{Explaining the Dimensionality of the Bottleneck Layer ($\MBS$)}
{Despite the notable success of DL in diverse wireless applications, there remains a gap in the availability of theoretical and systematic methods for analyzing DNNs. Specifically, there is an unresolved issue concerning the optimal dimensionality of the bottleneck layer in autoencoders, resulting in a lack of standardized approaches for automatically selecting this dimension. This section presents an information-theoretic methodology aimed at understanding the learning dynamics and determining the optimal dimensionality of the bottleneck layer, i.e. $\MBS$, in the proposed autoencoder. Our emphasis is on the crucial role of mutual information (MI) in quantifying learning from data. We utilize this methodology to investigate the layer-wise information flow and the inherent dimensionality of the bottleneck layer ($\MBS$) employing the data processing inequality and identifying a bifurcation point in the information plane, respectively. These findings have direct implications for optimizing autoencoder design. In particular, we estimate the necessary dimensionality to minimize the number of probing beams while preserving the performance of subsequent RF precoders.}

\subsection{Explainability with Entropy}

To identify the optimal phase shifts for the RF beamformer, and considering that we learn the RF beamformer shifts from a limited number of probing beams (i.e., $\MBS$), it is crucial that the probing codebook conveys valuable information to the RF beam prediction function across the entire environment.  Hence, there exists an adequate dimension $M_{D}$ of the data at the bottleneck layer between NN$_{\text{enc}}$ and NN$_{\text{dec}}$ modules that  the compressed data will \textit{not} provide  enough information for the regression network to learn when $\MBS<M_{D}$. 
\subsubsection{How Information Flows During Training} \label{sec:entropyprof}
The training process involves backpropagation and stochastic gradient descent. Both the feedforward and backward passes are unidirectional, depending solely on previous variables, forming a Markov chain \cite{shwartz2017opening}. The encoder and decoder both enforce entropy in the bottleneck layer, emphasizing the autoencoder's role in maximizing the entropy within this hidden layer.

The  successive representations in a DNN create a straightforward Markov chain, denoted as $\mbf{X}\rightarrow \mbf{R}_1 \rightarrow \mbf{R}_2 \rightarrow \dots \rightarrow \mbf{R}_{{L_H}-1}\rightarrow \mbf{R}_{L_H}$, such that $I(\mbf{X}; \mbf{R}_1) \geq I(\mbf{X}; \mbf{R}_2) \geq \dots \geq I(\mbf{X}; \mbf{R}_{L_H})$, where $\mbf{X}$ is the input matrix, and $\mbf{R}_1, \mbf{R}_2, \dots, \mbf{R}_{L_H}$ represent successive hidden layer representations from the initial hidden layer to the output layer. Consequently, fundamental data processing inequalities (DPIs) exist in any feedforward DNN with $L_H$ hidden layers. Given the Auto-HP architecture in Fig.\ref{fig:NetArch}, we focus on the partial feedforward Markov chain between the input and the bottleneck layer, along with the partial dual Markov chain between the output and the bottleneck layer. DPIs are evident in both the encoder and decoder, i.e., $I(\widehat{\mbf{h}};   \mbf{r}_{\mathrm{RSSI}}) \geq  I(\widehat{\mbf{h}}; \mbf{y}_{\mathrm{RSSI}}),$ and $I(\boldsymbol{\theta}_b; \mbf{d}_1) \geq I(\boldsymbol{\theta}_b; \mbf{d}_2) \geq I(\boldsymbol{\theta}_b; \mbf{d}_3) \geq I(\boldsymbol{\theta}_b; \mbf{y}_{\mathrm{RSSI}})$, 
 where $  \mbf{r}$, $\mbf{d}_1$, $\mbf{d}_2$, $\mbf{d}_3$ are the hidden layer representations of the neural networks NN$_{\text{enc}}$ and NN$_{\text{dec}}$, as shown in Fig. \ref{fig:NetArch}.
 
 In case of no compression, $ \MBS\geq\NBS$, there also exists a second type of DPI associated with the layer-wise MI with target outputs’ network, i.e., $I(\boldsymbol{\theta}_b; \boldsymbol{\theta}_b^{\star})\geq I(\mbf{d}_1; \mbf{d}_1^{\star}) \geq I(\mbf{d}_2; \mbf{d}_2^{\star}) \geq I(\mbf{d}_3; \mbf{d}_3^{\star}) \geq H( \mbf{y}_{\mathrm{RSSI}}),$
where $\boldsymbol{\theta}^{\star}$ represents the optimal output phase (target) obtained without compression, and $\mbf{d}_1^{\star}$, $\mbf{d}_2^{\star}$ and $\mbf{d}_3^{\star}$ represent the successive hidden layers representations for NN$_{\text{dec}}$ that obtained  $\boldsymbol{\theta}^{\star}$ at the output layer.{ The \textbf{target phase} shift $\boldsymbol{\theta}_b^{\star}$ is derived without compressing the bottleneck layer, where the dimension of this layer matches the number of antennas $M_{BS}=N_{BS}$. Our goal is to achieve this phase shift by compressing the bottleneck layer as much as possible. It is important to note that having the bottleneck layer dimension equal to or larger than the number of antennas can be resource-intensive and impractical in many scenarios. Due to the significant overhead from the RSSI beams, the bottleneck dimension determines the number of RSSI beams during initial access. Therefore, our goal is to compress the bottleneck layer as much as possible while still accurately matching the target phase shift $\boldsymbol{\theta}_b^{\star}$. This is achieved by finding  $\MBS$  that maximizes entropy at the bottleneck, i.e.,   
\begin{equation}\label{equ:explainEntropy}
    I(\boldsymbol{\theta}_b; \boldsymbol{\theta}_b^{\star}) \approx H( \mbf{y}_{\mathrm{RSSI}}).
\end{equation}}

\subsubsection{Estimation of Mutual Information from Mini-Batches}


Suppose $\mbf{X}$ and $\mbf{Y}$ are two random variables with a joint probability density function (PDF) denoted as $p(\mbf{x},\mbf{y})$, and marginals given by $p(\mbf{x})$ and $p(\mbf{y})$. The Renyi's entropy of order $\alpha$ for $\mbf{X}$ is defined as
\begin{equation}\label{equ:entr}
{H_\alpha }({\mathbf{X}}) = \frac{1}{{1 - \alpha }}\log \int_{\mathbb{R}} {{p^\alpha }({\mathbf{x}})dx},
\end{equation}
where $ \alpha\geq 0$ and $\alpha\neq 1$. When $\alpha\rightarrow 1$, Renyi's entropy converges to Shannon’s differential entropy. The MI $I(\mbf{X}; \mbf{Y}) $ is then defined as the relative entropy between the joint distribution and the product distribution $p(\mbf{x})p(\mbf{y})$, and can be represented as
\begin{equation}\label{equ:MI}
I({\mathbf{X}};{\mathbf{Y}}) = H({\mathbf{X}}) + H({\mathbf{Y}}) - H({\mathbf{X}},{\mathbf{Y}}),
\end{equation}
where $H(\mbf{X})$ and $H(\mbf{Y})$ represent the marginal entropies of $\mbf{X}$ and $\mbf{Y}$, while $H({\mathbf{X}},{\mathbf{Y}})$ denotes their joint entropy. In the context of DNN, it is not feasible to analytically evaluate the MI in \eqref{equ:MI}, as \eqref{equ:entr} necessitates the precise estimation of the PDF of $\mbf{X}$ and $\mbf{Y}$ in a high-dimensional space. Consequently, the value of MI must be estimated from a limited number of samples, such as mini-batches.

To compute Renyi's entropy and MI, we follow the methodology outlined in \cite{Sanchez2015,theoauto2021,yu2019understanding}. We begin with defining a batch ${\mathbf{X}} = \left\{ {{{\mathbf{x}}_i}} \right\}_{i = 1}^{N}$, comprising an independent and identically distributed (iid) sample of $N$ realizations of $\mbf{X}$.  Subsequently, we define a matrix-based counterpart to Rényi’s $\alpha$-entropy \eqref{equ:entr} for a matrix $\mbf{A}\in \mathbb{R}^{n\times n}$ with $\mathrm{trace}(\mbf{A}) = 1$, as follows
\begin{equation}\label{equ:S_a}
{S_\alpha }({\mathbf{A}}) = \frac{1}{{1 - \alpha }}\log \left[ {\sum\nolimits_{i = 1}^n {{\lambda _i}{{({\mathbf{A}})}^\alpha }} } \right],
\end{equation}
where ${{\mathbf{A}}_{ij}} = \frac{1}{n}\frac{{{{\mathbf{K}}_{ij}}}}{{\sqrt {{{\mathbf{K}}_{ii}}{{\mathbf{K}}_{jj}}} }}$ such that  $\mbf{K}$ is a Gram matrix derived by evaluating a real-valued positive definite kernel $\kappa: \mathbb{R} \times \mathbb{R} \rightarrow \mathbb{R}$ on all pairs of data points, yielding $(\mbf{K})_{ij} = \kappa(\mbf{x}_i,\mbf{x}_j)$. $\lambda_i{(\mbf{A})}$ represents the $\nth{i}$ eigenvalue of $\mbf{A}$. 
Therefore, a matrix-based estimation of the joint entropy  can be computed as 
\begin{equation}
{S_\alpha }({\mathbf{A}},{\mathbf{B}}) = {S_\alpha }\left( {\frac{{{\mathbf{A}} \odot {\mathbf{B}}}}{{\mathrm{trace}({\mathbf{A}} \odot {\mathbf{B}})}}} \right),
\end{equation}
where $\odot$ represents the Hadamard product, and the matrix $\mbf{B}$ is derived similarly to $\mbf{A}$ but is based on samples ${\mathbf{Y}} = \left\{ {{{\mathbf{y}}_i}} \right\}_{i = 1}^N$ from the corresponding layer of interest obtained from the same realizations of $\mbf{X}$. It's noteworthy that matrices $\mbf{A}$ and $\mbf{B}$ simplify the estimation of the joint distribution through pairwise element multiplication. Ultimately, the matrix-based Rényi’s MI is expressed, in analogy with Shannon’s definition \eqref{equ:MI}, as:
\begin{equation}\label{equ:I_AB1}
{I_\alpha }({\mathbf{A}};{\mathbf{B}}) = {S_\alpha }({\mathbf{A}}) + {S_\alpha }({\mathbf{B}}) - {S_\alpha }({\mathbf{A}},{\mathbf{B}}).
\end{equation}
{Hence, to calculate the mutual information \(I(\boldsymbol{\theta}_b; \boldsymbol{\theta}_b^{\star})\) using Rényi's entropy, can be expressed as:
\begin{equation}\label{equ:I_AB}
I(\boldsymbol{\theta}_b; \boldsymbol{\theta}_b^{\star}) = S_\alpha(\boldsymbol{\theta}_b) + S_\alpha(\boldsymbol{\theta}_b^{\star}) - S_\alpha(\boldsymbol{\theta}_b, \boldsymbol{\theta}_b^{\star}),
\end{equation}
where \(\boldsymbol{\theta}_b\) and \(\boldsymbol{\theta}_b^{\star}\) have the same dimension of $\NBS \times 1$.}

	\begin{algorithm}[t!]
		\caption{ \textsc{\small{Bisection Search of Bottleneck Dimension}}}\label{Alg:2}\footnotesize
			\begin{algorithmic}[1]
				\State {\textbf{Initialize}: $\mbf{h}_i \in \mathcal{H}_U \forall i \in [1, N_D N_U]$, $\NBS$ \label{Alg2:line:1}}
                \State $L \leftarrow 0$, $D \leftarrow \NBS-1$
                \While{$L \leq D$}
                    \State $\MBS \leftarrow \lceil(L+D)/2 \rceil  $
                    \State $R \gets \textsc{ Entropy Condition Check  } (\mbf{h}_u, \forall u, \MBS) $
                        \If {$R$}
                        \State $D\gets \MBS -1$
                        \Else
                        \State $L\gets \MBS+1$
                        \EndIf
                \EndWhile\\
            \Return $\MBS$
            \vspace{2pt}
            \hrule 
            \vspace{2pt}
				
				\Procedure{\small{Entropy Condition Check }}{$\mbf{h}_u, \forall u, K$}
                        \State $\mathcal{H}^{\text{train}}_{U},\mathcal{H}^{\text{valid}}_{U}\gets \mathcal{H}_{U}$
                        \State NN$_{\text{Auto-HP}}\gets$ construct NN with $(\MBS)$ bottleneck dimension
                        \State $N_{B}\gets length{\mathcal{H}^{\text{train}}_{U}}/B $
                        \For{$n_e = 0:  N_{Epochs}$}
                            \For{$n_b = 0: N_{B}$}
                            \State $\mbf{H}^{n_b}_{U} \gets getBatch({\mathcal{H}^{\text{train}}_{U}},n_b)$
                            \State $\boldsymbol{\Theta}_{n_b}, \mbf{Y}_{\mathrm{RSSI,n_b}} \gets $ NN$_{\text{Auto-HP}}.train(\mbf{H}^{n_b}_{U})$
                            \State $S(\mbf{Y}_{\mathrm{RSSI},n_b}) \gets$ as per \eqref{equ:S_a} 
                            \State $I(\boldsymbol{\Theta}_{n_b};\boldsymbol{\Theta}_{n_b}^{\star}) \gets$ as per  \eqref{equ:I_AB}
                            \EndFor
                            \State {\small $\bar{I}_{\boldsymbol{\Theta}},\bar{S}_{\mbf{Y}_{\mathrm{RSSI}}}\gets Avg(I(\boldsymbol{\Theta}_{n_b};\boldsymbol{\Theta}_{n_b}^{\star}),S(\mbf{Y}_{\mathrm{RSSI},n_b}), N_{B})$}
                            \If {$\bar{S}_{\mbf{Y}_{\mathrm{RSSI}}} \simeq k\times\bar{I}_{\boldsymbol{\Theta}}$}
                            \State \textbf{break and return }$1$
                            \ElsIf {$earlyStop$}
                            \State \textbf{break and return }$0$
                            \EndIf
                        \EndFor\\
				\Return $0$
				\EndProcedure
			\end{algorithmic}
	\end{algorithm}
\subsubsection{Determining Adequate Bottleneck Dimension}
 To determine the adequate number of the probing beams $\MBS$ which is corresponding to the bottelneck dimension of NN$_{\text{Auto-HP}}$, we propose a bisection-based algorithm based only on the evolution of $H(\mbf{y}_{\mathrm{RSSI}})$. Our proposed methodology  involves finding $\MBS$ within an array featuring a range of effective search dimensions. The lower bound is confined to 0, while the upper bound is determined by the number of phase shifts denoted by the final layer dimension of NN$_{\text{Auto-HP}}$, denoted as $\NBS$. {In particular, the algorithm performs a binary search between $L = 0$ and $D = N_{BS} - 1$. The number of iterations for a binary search is typically $\mathcal{O}(\log N_{BS})$.}  The algorithm consists of searching the sorted array by repeatedly dividing the search interval in half and check if the entropy condition holds, i.e. $ H(\mbf{y}_{\mathrm{RSSI}}) \simeq k \times I(\boldsymbol{\theta}_b,\boldsymbol{\theta}_b^{\star})$ holds at each array division when training the NN$_{\text{Auto-HP}}$ (Algorithm \ref{Alg:2}), where $k$ represents the approximation level. We have provided discussion on why $H(\mbf{y}_{\mathrm{RSSI}})$ holds as a performance metric in previous section \ref{sec:entropyprof}.
 \begin{figure}
     \centering
     \includegraphics[scale=0.25]{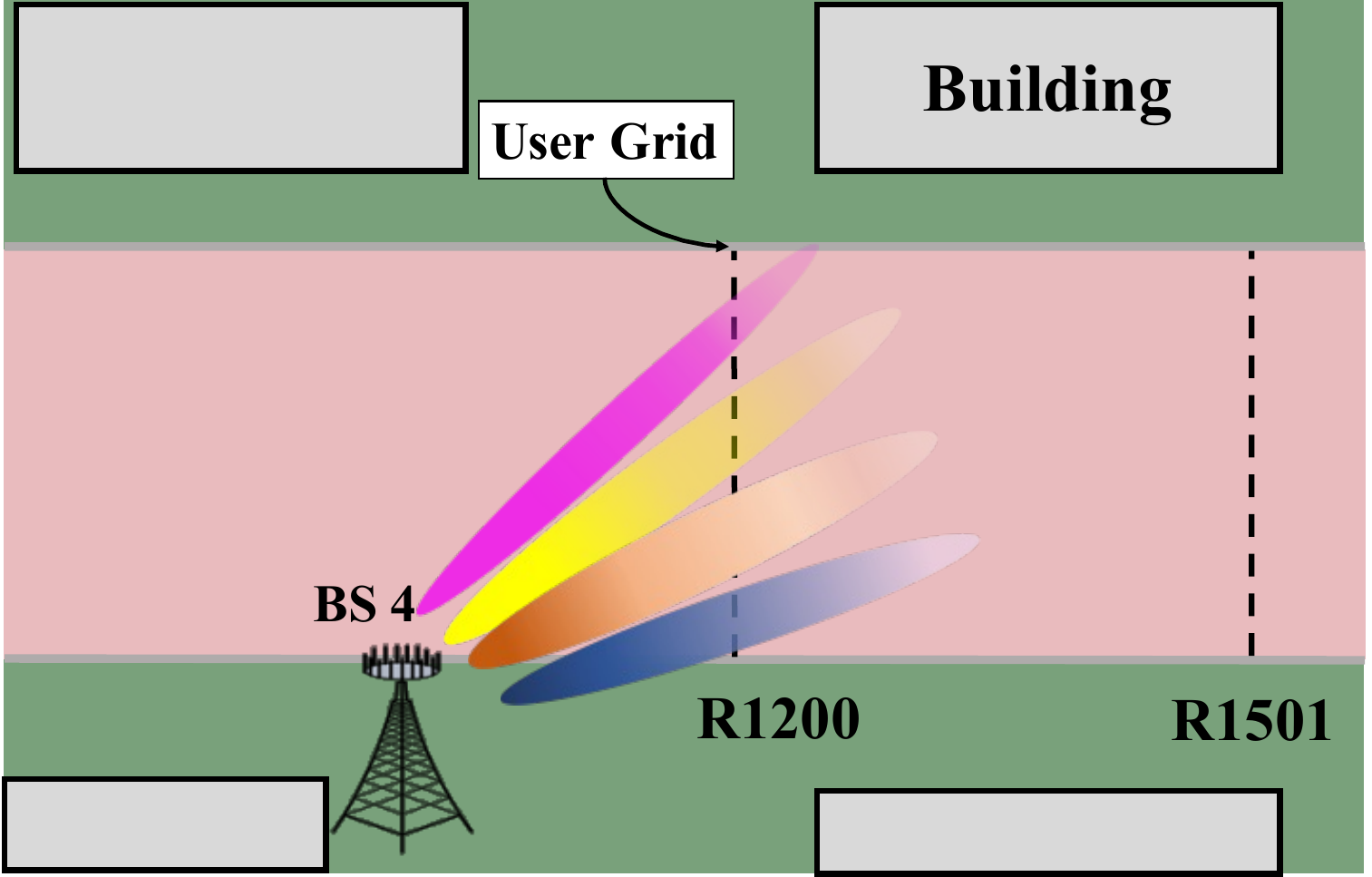}
			\caption{Illustration of the area covered in the DeepMIMO channel datasets by considered communication scenario.}\label{fig:DeepMIMO}
 \end{figure}
 \begin{figure}
     \centering
     \includegraphics[width=0.82\columnwidth]{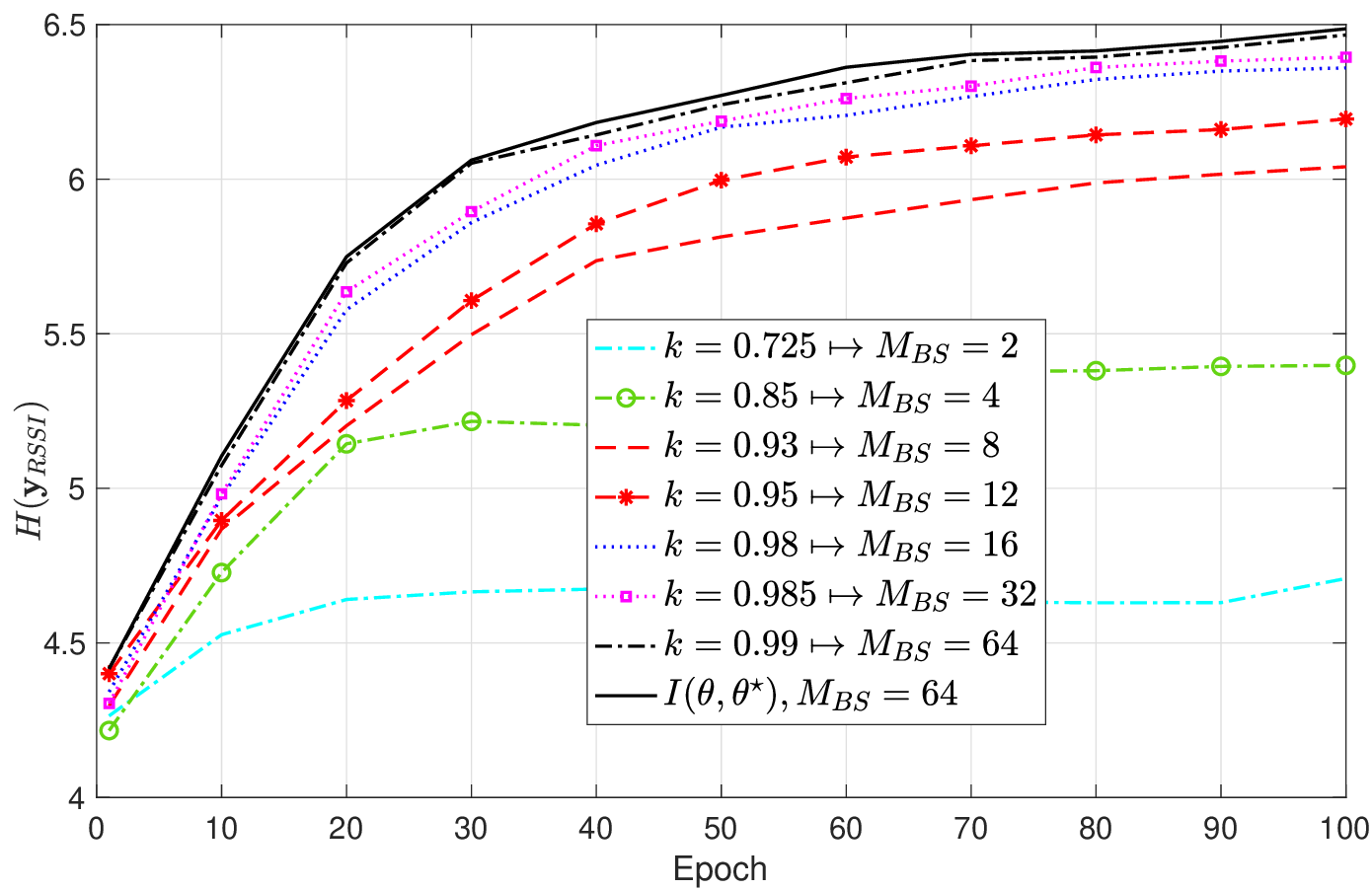}
			\caption{Evolution of $H(\mbf{y}_{\mathrm{RSSI}})$ versus number of epochs for different bottleneck dimension $\MBS$. }\label{fig:MI_epoch}
 \end{figure}
 
  \begin{figure}[t]
     \centering
     \includegraphics[width=0.85\columnwidth]{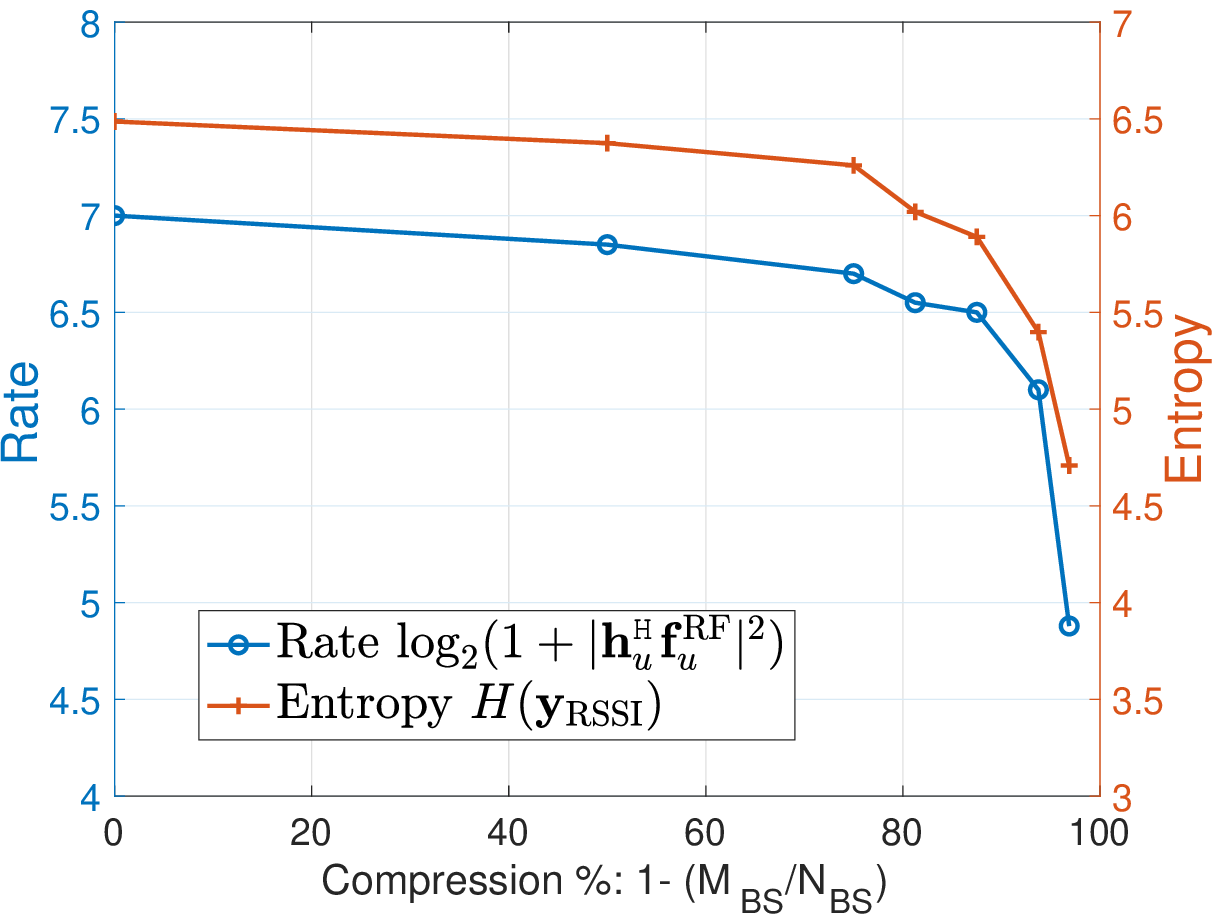}
			\caption{Rate $\log_2(1+{|\mathbf{h}_u^{\mathsf{H}}{\mathbf{f}^{\mathrm{RF}}_u}|^2})$ and entropy $H(\mbf{y}_{\mathrm{RSSI}})$ versus compression rate ($1-\frac{\MBS}{\NBS}$).}\label{fig:rate_compress}
 \end{figure}
 
In \textsc{ Entropy Condition Check  } procedure from Algorithm \ref{Alg:2}, we have a dataset of channel realizations $\mathcal{H}_U$ that is divided into training and validation datasets ($\mathcal{H}^{\text{train}}_{U},\mathcal{H}^{\text{valid}}_{U}$), where the training set is also divided into $N_{B}$ mini-batches.  We then  use a mini-batch of channel realizations, $\mbf{H}^{n_b}_{U}=\left\{ {{{\mathbf{h}_{}}_i}} \right\}_{i = 1}^{B}$ where $B$ is the batch size, to train a new NN$_{\text{Auto-HP}}$ with bottleneck layer size given by the dimension $\MBS$ as the middle position of the searching array. We define ${\mathbf{Y}_{\mathrm{RSSI},n_b}} = \left\{ {{{\mathbf{y}_{\mathrm{RSSI},i}}}} \right\}_{i = 1}^{B}$ and ${\boldsymbol{\Theta}_{n_b}} = \left\{ {\boldsymbol{\theta}_{i}} \right\}_{i = 1}^{B}$.  During each training iteration, the updated NN$_{\text{Auto-HP}}$ is employed to propagate the data through each layer and estimate MI between the target phase shift ${\boldsymbol{\Theta}_{n_b}^\star}$ and output data $I({\boldsymbol{\Theta}_{n_b}} ,\boldsymbol{\Theta}_{n_b}^{\star})$ using \eqref{equ:I_AB}, and the entropy of codes $S(\mbf{Y}_{\mathrm{RSSI},n_b})$ using \eqref{equ:S_a}. Moreover, $I({\boldsymbol{\Theta}_{n_b}} ,{\boldsymbol{\Theta}_{n_b}^\star})$ is rounded to two decimals, and the condition $S(\mbf{Y}_{\mathrm{RSSI},n_b}) \simeq k \times I({\boldsymbol{\Theta}_{n_b}} ,{\boldsymbol{\Theta}_{n_b}^\star})$ is checked at the end of each epoch. s these metrics are computed at the mini-batch level, to expedite computation, MI and entropy are estimated every 10 mini-batches and averaged at the end of each epoch. It is worth noting that these procedures are excluded from Algorithm \ref{Alg:2} to maintain clarity. Subsequently, if the condition is satisfied, the interval is narrowed to the lower half; otherwise, it is narrowed to the upper half. This iterative process continues until the interval becomes empty, resulting in the determination of $\MBS$. 

\begin{figure*}
    \begin{subfigure}[b]{0.3\textwidth}
			\includegraphics[width=0.9\columnwidth]{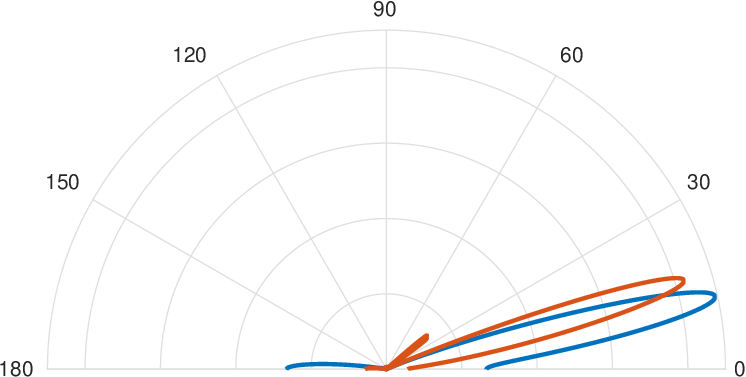}
			\caption{}\label{fig:Prob2}
    \end{subfigure} \kern-1em
    \begin{subfigure}[b]{0.3 \textwidth}
\includegraphics[width=0.9\columnwidth]{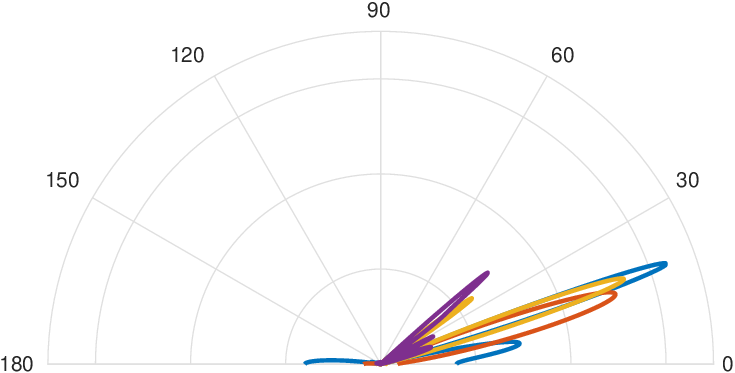}
			\caption{}\label{fig:Prob4}
    \end{subfigure}
    \begin{subfigure}[b]{0.3 \textwidth}
\includegraphics[width=0.9\columnwidth]{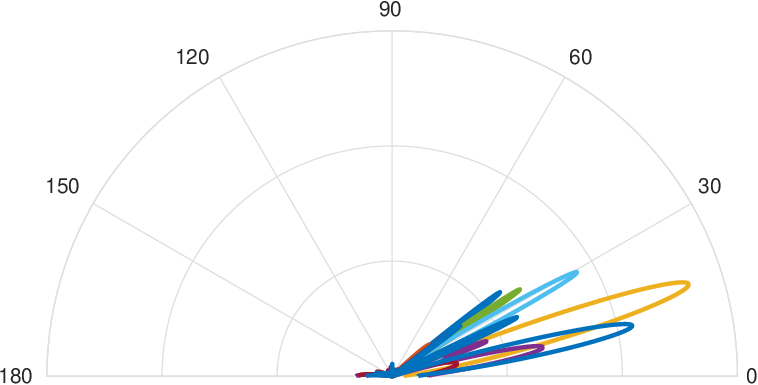}
			\caption{}\label{fig:Prob8}
    \end{subfigure} \kern-1em
    \quad\begin{subfigure}[b]{0.3\textwidth}
			\includegraphics[width=0.9\columnwidth]{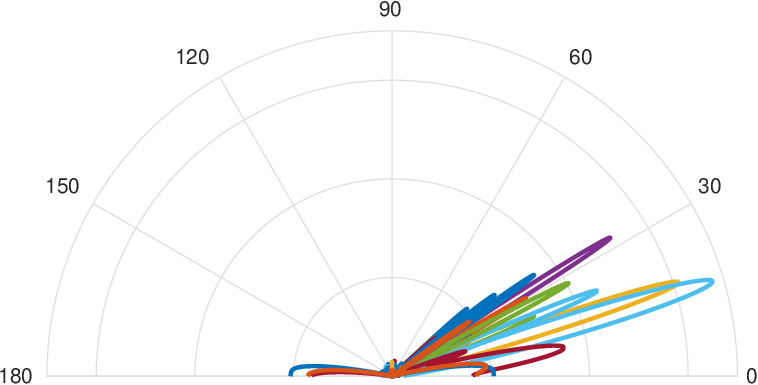}
			\caption{}\label{fig:Prob16}
    \end{subfigure}\quad
    \begin{subfigure}[b]{0.3 \textwidth}
\includegraphics[width=0.9\columnwidth]{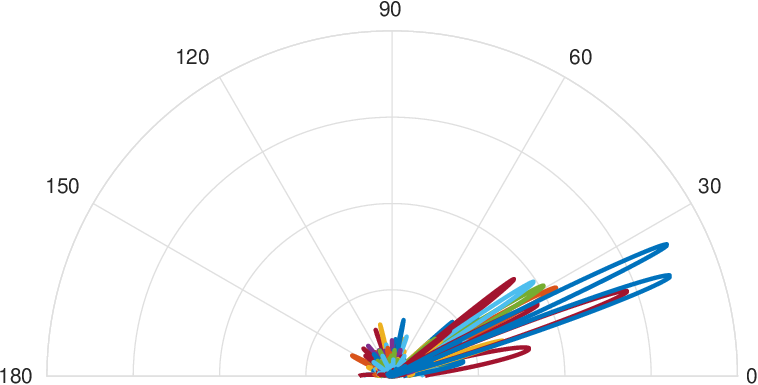}
			\caption{}\label{fig:Prob32}
    \end{subfigure}\qquad
    \begin{subfigure}[b]{0.3 \textwidth}
\includegraphics[width=0.9\columnwidth]{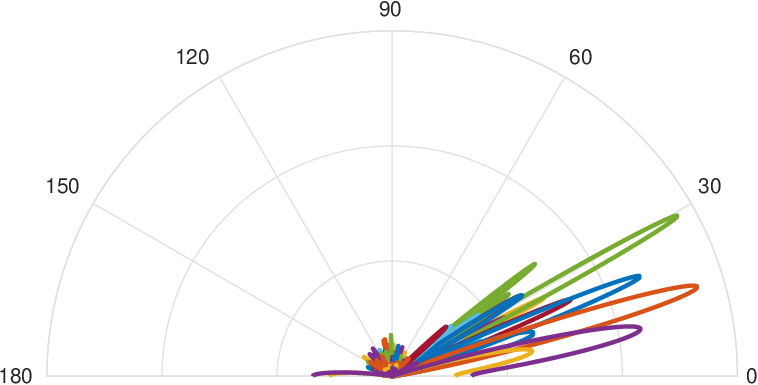}
			\caption{}\label{fig:Prob64}
    \end{subfigure}
    \caption{ Radiation patterns of learned probing beams in $\mbf{P}$ for: (a) $\MBS=2$, (b) $\MBS=4$, (c) $\MBS=8$, (d) $\MBS=16$, (e) $\MBS=32$, and (f) $\MBS=64$. }
    \label{fig:Prob}
\end{figure*}

\begin{figure*}[t]
    \begin{subfigure}[b]{0.32\textwidth}
			\includegraphics[width=0.99\columnwidth]{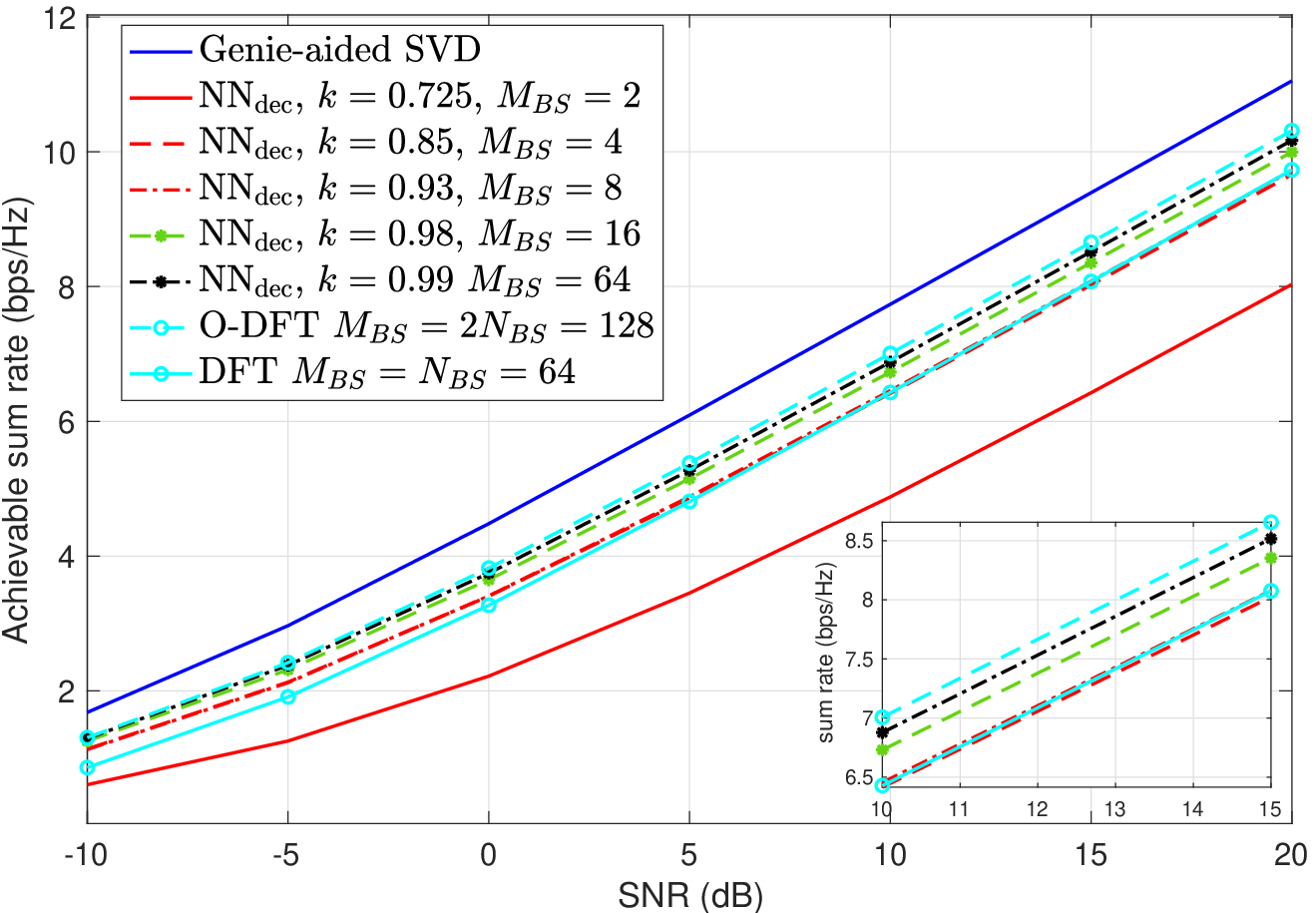}
			\caption{}\label{fig:SRU1}
    \end{subfigure}
    \begin{subfigure}[b]{0.32 \textwidth}
\includegraphics[width=0.99\columnwidth]{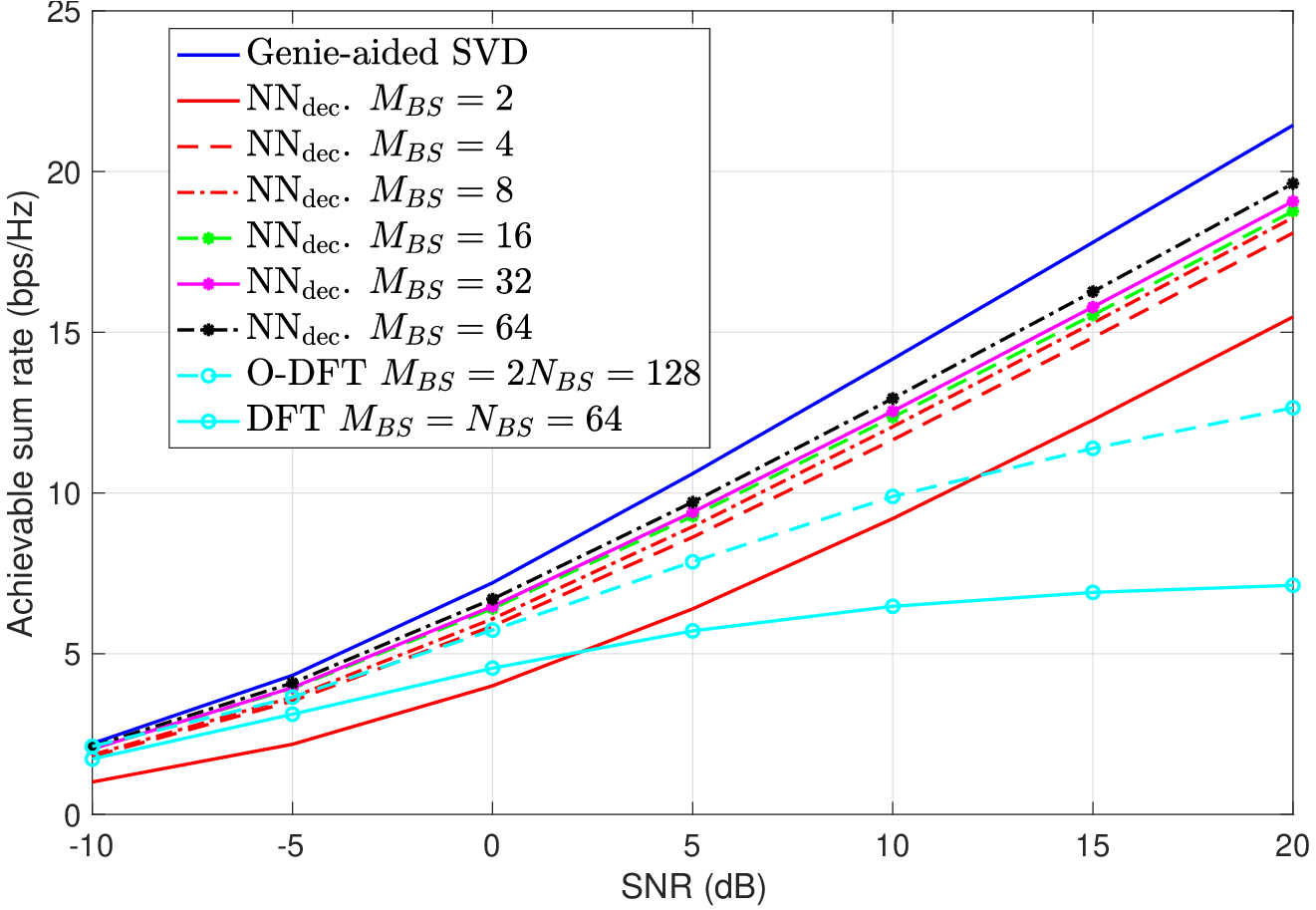}
			\caption{}\label{fig:SRU2}
    \end{subfigure}
    \begin{subfigure}[b]{0.32 \textwidth}
\includegraphics[width=0.99\columnwidth]{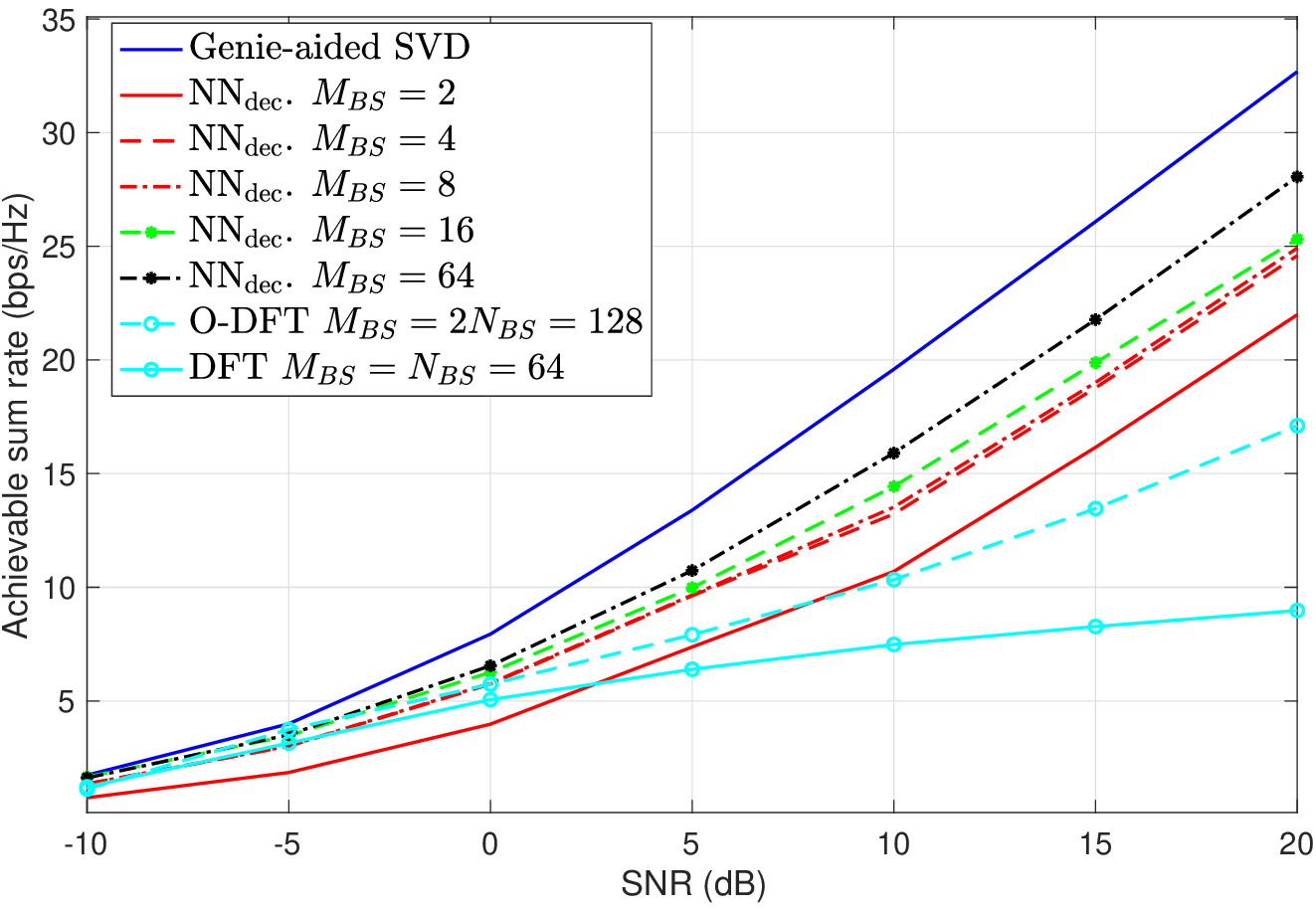}
			\caption{}\label{fig:SRU4}
    \end{subfigure}
    \caption{ The achievable sum rates versus SNR for single and multi-users cases: (a) $N_U=1$ , (b) $N_U=2$, and (c) $N_U=4$. }
    \label{fig:SRU}
\end{figure*}

\begin{figure}[t]\centering
    \begin{subfigure}[b]{0.4\textwidth}
			\includegraphics[width=0.85\columnwidth]{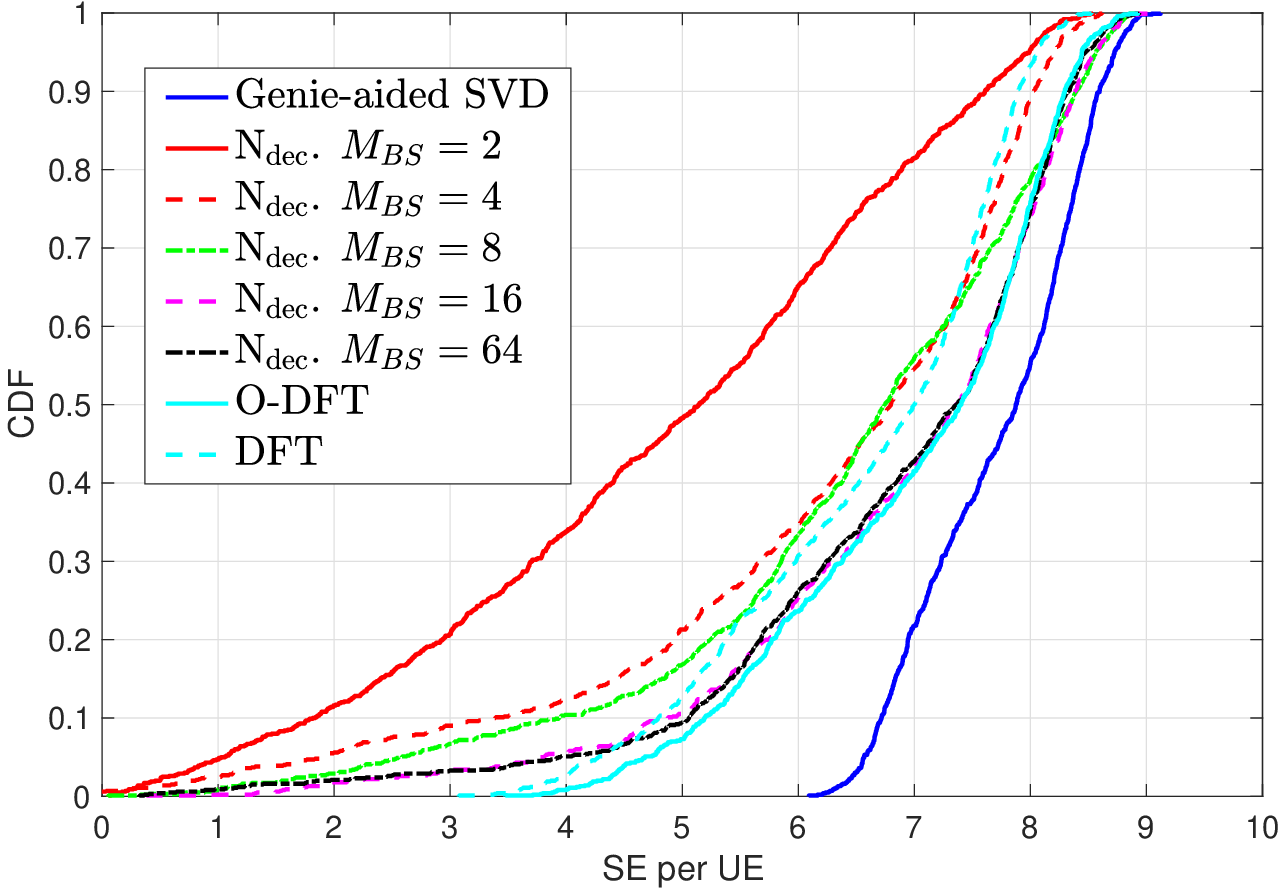}
			\caption{}\label{fig:CDFU1}
    \end{subfigure}
    \begin{subfigure}[b]{0.4 \textwidth}
\includegraphics[width=0.85\columnwidth]{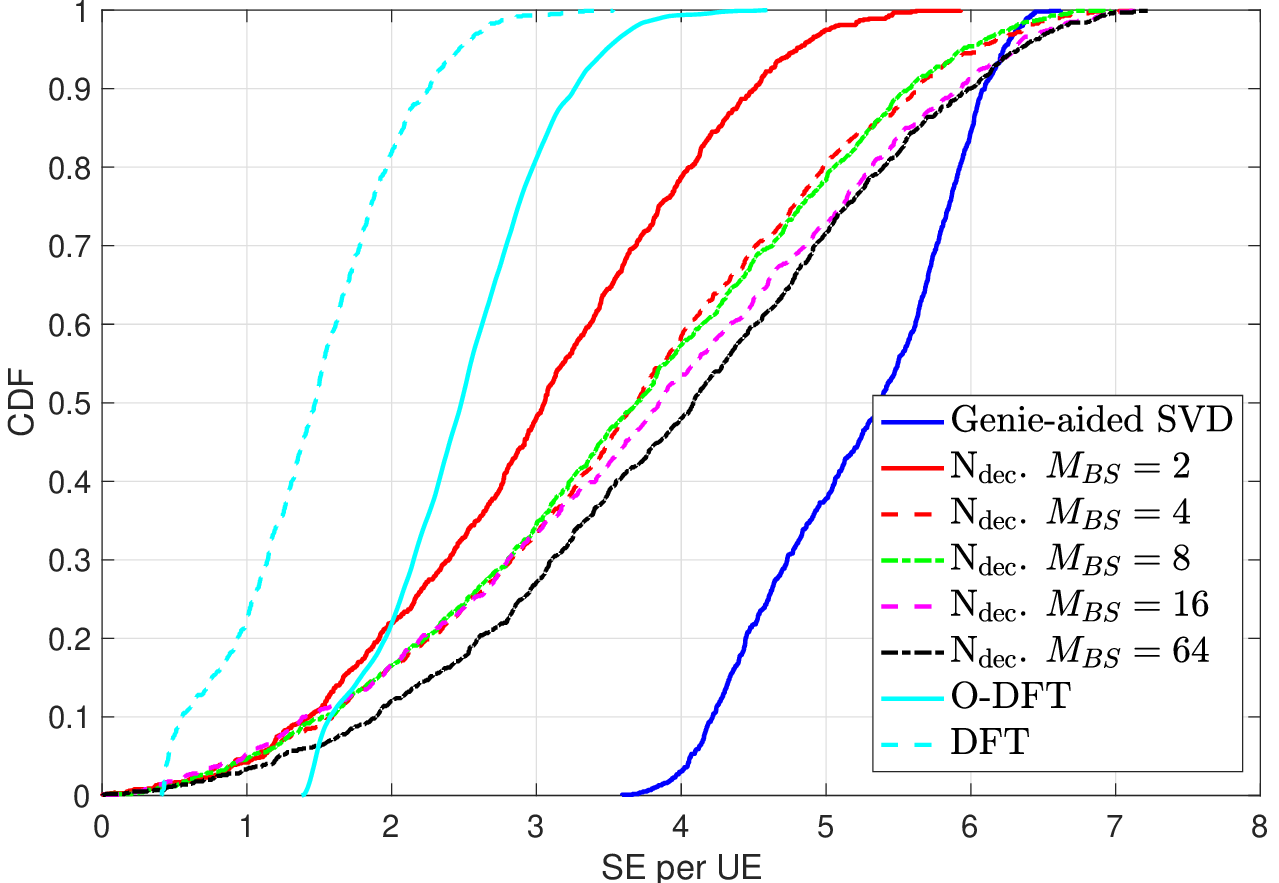}
			\caption{}\label{fig:CDFU4}
    \end{subfigure}
    \caption{ Cumulative distribution functions (CDF) of the achievable rates per UE for single and multi-users cases at {SNR $=\unit[10]{dB}$}: (a) $N_U=1$, and (b) $N_U=4$.}
    \label{fig:CDFU}
\end{figure}
\section{Simulation Results}
In this section, we evaluate the performance of the proposed E2E learning-based multi-UE hybrid precoding design approach using realistic 3D ray-tracing simulations.
\subsection{Dataset Generation}

{We generate our dataset using the publicly available DeepMIMO dataset \cite{Alkhateeb2019}. As depicted in Fig. \ref{fig:DeepMIMO}, our analysis focuses on BS 4 in the street-level outdoor scenario labeled 'O1-28', which communicates with $N_D=16,000$ UEs spanning from row R1200 to R1500, operating at a carrier frequency of $\unit[28]{GHz}$. {The BS is equipped with $\NBS=64$ antennas and $\NRF=4$ RF chains.} For each UE, we initially generate the channel vector using the DeepMIMO dataset generator. Subsequently, random noise is introduced to the channel matrix, with the noise power computed based on a bandwidth of $\unit[0.5]{GHz}$ and a receive noise figure of $\unit[5]{dB}$. The dataset created is then utilized to train our proposed auto-precoder neural network, incorporating a b-bit quantizer layer ($Q_b(.)$) with $b=3$ and employing the customized loss function detailed in Section \ref{sec:loss}.  The dataset is partitioned into $90\%$ and $10\%$ for the training and validation set, respectively.} 

{We use the Adam optimizer, a batch size of $B=128$, a learning rate of  0.004, and $N_{Epochs}=100$ epochs. We utilized the normalized radial basis function (RBF) kernel, $\kappa(\mathbf{x}_i, \mathbf{x}_j; \sigma) = \exp \left( \frac{1}{2\sigma^2} \|\mathbf{x}_i - \mathbf{x}_j\|^2 \right).$ The kernel's bandwidth parameter $\sigma$ is crucial as it significantly affects the resulting estimates. To determine $\sigma$, we applied Silverman’s rule of thumb for Gaussian density estimation, which is given by $\hat{\sigma} = h n^{-1/(4+d)}$, where $h$ represents the standard deviation of the samples, $n$ denotes the sample size, and $d$ is the sample dimensionality.}

\subsection{Explainability through Mutual Information}

Fig. \ref{fig:MI_epoch} shows the evolution of the MI of the bottelneck layer for different $\MBS$ and $k$ values across the training epochs {where $k$ represents the approximation level that holds $ H(\mbf{y}_{\mathrm{RSSI}}) \simeq k\times I(\boldsymbol{\theta}_b,\boldsymbol{\theta}_b^{\star})$.}
When minimizing the loss function and following Algorithm \ref{Alg:2}, we check during training whether the condition $ H(\mbf{y}_{\mathrm{RSSI}}) \simeq k \times  I(\boldsymbol{\theta}_b,\boldsymbol{\theta}_b^{\star})$ holds for the obtained $\MBS$ such that $\MBS<\NBS$ or the condition does not hold, i.e., $ H(\mbf{y}_{\mathrm{RSSI}}) < k \times I(\boldsymbol{\theta}_b,\boldsymbol{\theta}_b^{\star})$. It is worth noting that $\boldsymbol{\theta}_b^{\star}$ is obtained when training the model without compression that is when $\MBS=\NBS=64$. As shown in Fig. \ref{fig:MI_epoch}, There is a degradation in  performance of the regression network when training for a   low values of $\MBS=[2,4]$. For $k=0,93$,  the lowest dimension $\MBS=8$ that holds $ H(\mbf{y}_{\mathrm{RSSI}}) \simeq 0.93\times I(\boldsymbol{\theta}_b,\boldsymbol{\theta}_b^{\star})$, that is, the maximum compression of data that retain an approximately similar information when having no compression. Results show that the entropy of received probing signals $H(\mbf{y}_{\mathrm{RSSI}})$ increases as the bottleneck layer dimension  increases, with the upper bound being the entropy $I(\boldsymbol{\theta}_b; \boldsymbol{\theta}_b^{\star})$ when the autoencoder is trained for $\MBS=\NBS=64$.

Figure \ref{fig:rate_compress} extends the investigation into the impact of compression rate on both the achievable rate and the entropy of the bottleneck layer. Furthermore, it confirms the conclusions drawn from the preceding figure, indicating a significant decrease in entropy and rates for $\MBS=[2,4]$, while for $\MBS\geq 8$, the drop is approximately less than 7$\%$.
 
 \subsection{Probing Beam Patterns}

 To gain deeper insights into the performance of the probing beam, we analyze the acquired beam patterns. For enhanced visual comprehension, the BS is deployed with uniform linear arrays (ULA) that  beamform in the azimuth domain. With just 2 and 4 probing beams, it can be seen that small area is covered. However, with more than 8 probing beams, the probing beam patterns are meticulously tailored to the propagation environment, notably directing a significant portion of energy towards the reflectors positioned on the right side of the BS covering more areas, as depicted in Figure \ref{fig:Prob}. These probing beams often exhibit multiple lobes, a feature likely facilitating more effective capture of channel information. As the number of probing beams increases from 2 to 64, the acquired probing beams exhibit significantly broader spatial coverage, as illustrated in Figure \ref{fig:Prob}. Leveraging the augmented information acquired through the probing beam pairs, the  RF precoding beam generator NN$_{\text{dec}}$ also adeptly learns to concentrate energy with better accuracy. {It is worth noting that our proposed end-to-end solution inherently takes advantage of the channel sparsity. The sparsity of the channels is a key factor that allows us to design a limited number of sensing beams effectively. }

\subsection{Single-User vs. Multi-User Achievable Rates}

To study the achievable rates of the single and multi-user cases, we provide plots on achievable sum rates versus SNR in Fig.\ref{fig:SRU} and cumulative distribution functions (CDF) for the achievable rates per UE in Fig. \ref{fig:CDFU}. We consider the singular value decomposition (SVD)-based beamforming as a benchmark scheme as it achieves the maximal achievable rate \cite{Heath2016OverviewmmWave} with perfect CSI knowledge. Note that the SVD upper bound can only be attained when the channel of each UE is known, and the phase shifters are not quantized. {Additionally, we benchmark the proposed E2E learning approach with RF precoders designed based on both the DFT-based codebook scanning directions with $\NBS$ candidate beams at the BS \cite{li2019deep, Andrews2022BF} and O-DFT codebook \cite{Andrews2022BF}, which utilizes $2\NBS$ oversampled DFT beams. In both cases, we employ ZF as the BB precoder.}
In Figure \ref{fig:SRU1}, we plot the achievable sum rate for a single UE case  following equ. \eqref{eq:rate} without the interference term in the denominator, it is shown that the $\MBS=16$ provides a good performance outperforming the oversampled DFT-based codebook. In Figures \ref{fig:SRU2} and \ref{fig:SRU4}, we plot the achievable sum rates for a multi-user case  following equ. \eqref{eq:rate} for $N_U=2$ and $N_U=4$, respectively. {It is worth noting that the DFT-based and O-DFT-based precoders are limited to a set of predefined beams within their codebook. This restriction inherently limits their ability to adapt to the dynamic and complex multi-user environments, leading to suboptimal beamforming and higher interference among users.} {It is worth noting that for our simulated region, we require 32 DFT beams out of the 64 DFT beams, and for the O-DFT, we need 64 beams out of the 128 O-DFT beams. Our simulation results demonstrate that our proposed solution with $\MBS=8$ provides good performance, outperforming both the DFT-based and oversampled DFT-based codebooks. Consequently, the beam training overhead is reduced by $75\%$ compared to DFT and $87.5\%$ compared to O-DFT. }
As the number of probing beams increase $\MBS$, the performance increases as more probing beams are able to collect more information about the UEs in the system and provides more accurate RF precoders. 
In Fig. \ref{fig:CDFU1}, Fig. \ref{fig:CDFU4} present the cumulative distribution function (CDF) of the achievable rate obtained for the considered codebook designs for $N_U=1$ and $N_U=4$, respectively. From the plot, it can be observed that the distribution of the achievable rate is superior for the codebook designed using the proposed  learning-based approach as compared to the DFT codebook for a smaller number of $\MBS=8$ beams and to the O-DFT codebook for a number of $\MBS=16$ beams. Additionally, our proposed DRL solution achieves a better performance than that of DFT in terms of the $10\%$ percentile of the achievable rate with only $\MBS=8$ beams.

\section{Conclusion}
An explainable E2E learning-based framework has been proposed for a joint design of probing beam codebooks and the hybrid precoders for a multi-user mmWave system. The proposed framework achieves several manifolds. Firstly, it diverges from conventional methods that rely on generic or randomly chosen beam configurations by focusing on learning  a limited number of optimized probing vector which reduces latency. This approach effectively reduces beam training latency and allows for a more targeted concentration of measurement power in the most promising spatial directions, based on learning the user distribution and environmental conditions. Secondly, through advanced deep learning techniques, our model is capable of directly predicting hybrid beamforming vectors for multiple users based on compressed probing measurements,while mitigating interference between users. Lastly, using information-theoretic methodology  to maximize the entropy of the bottleneck layer, our system provides transparency in selecting the limited number of probing beams, which helps to demystify the black-box nature of deep learning and reduces uncertainty in predicting beamforming vectors. 


\begin{thebibliography}{10}
\providecommand{\url}[1]{#1}
\csname url@samestyle\endcsname
\providecommand{\newblock}{\relax}
\providecommand{\bibinfo}[2]{#2}
\providecommand{\BIBentrySTDinterwordspacing}{\spaceskip=0pt\relax}
\providecommand{\BIBentryALTinterwordstretchfactor}{4}
\providecommand{\BIBentryALTinterwordspacing}{\spaceskip=\fontdimen2\font plus
\BIBentryALTinterwordstretchfactor\fontdimen3\font minus \fontdimen4\font\relax}
\providecommand{\BIBforeignlanguage}[2]{{%
\expandafter\ifx\csname l@#1\endcsname\relax
\typeout{** WARNING: IEEEtran.bst: No hyphenation pattern has been}%
\typeout{** loaded for the language `#1'. Using the pattern for}%
\typeout{** the default language instead.}%
\else
\language=\csname l@#1\endcsname
\fi
#2}}
\providecommand{\BIBdecl}{\relax}
\BIBdecl

\bibitem{G2019WSR}
H.~Guo \emph{et~al.}, ``Weighted sum-rate maximization for intelligent reflecting surface enhanced wireless networks,'' in \emph{Proc. IEEE Global Conmun. Conf. (GLOBECOM)}, 2019, pp. 1--6.

\bibitem{Liu2021BF}
R.~Liu, M.~Li, Q.~Liu, and A.~L. Swindlehurst, ``Joint symbol-level precoding and reflecting designs for {IRS}-enhanced {MU-MISO} systems,'' \emph{{IEEE} Trans. Wireless Commun.}, vol.~20, no.~2, pp. 798--811, Feb. 2021.

\bibitem{Rehman2021BF}
H.~Ur~Rehman \emph{et~al.}, ``Joint active and passive beamforming design for {IRS}-assisted multi-user {MIMO} systems: A {VAMP}-based approach,'' \emph{{IEEE} Trans. Commun.}, vol.~69, no.~10, pp. 6734--6749, Oct. 2021.

\bibitem{Wang2021BF}
W.~Wang \emph{et~al.}, ``Joint beam training and positioning for intelligent reflecting surfaces assisted millimeter wave communications,'' \emph{{IEEE} Trans. Wireless Commun.}, vol.~20, no.~10, pp. 6282--6297, Oct. 2021.

\bibitem{Icc202BFRIS}
J.~Wang, Y.-C. Liang, S.~Han, and Y.~Pei, ``Robust beamforming and phase shift design for irs-enhanced multi-user {MISO} downlink communication,'' in \emph{Proc. IEEE Int. Conf. Commun. (ICC)}, Jun. 2020, pp. 1--6.

\bibitem{2019WSBF}
H.~Guo, Y.-C. Liang, J.~Chen, and E.~G. Larsson, ``Weighted sum-rate maximization for intelligent reflecting surface enhanced wireless networks,'' in \emph{Proc. IEEE Global Conmun. Conf. (GLOBECOM)}, Dec. 2019, pp. 1--6.

\bibitem{SMRIS2021WCNC}
Y.~Xiu, W.~Sun, J.~Wu, G.~Gui, N.~Wei, and Z.~Zhang, ``Sum-rate maximization in distributed intelligent reflecting surfaces-aided mmwave communications,'' in \emph{Proc. IEEE Wireless Commun. and Netw. Conf. (WCNC)}, Apr. 2021, pp. 1--6.

\bibitem{Gior2019BM}
M.~Giordani, M.~Polese, A.~Roy, D.~Castor, and M.~Zorzi, ``A tutorial on beam management for 3{GPP} {NR} at mm{W}ave frequencies,'' \emph{{IEEE} Commun. Surveys Tuts.}, vol.~21, no.~1, pp. 173--196, Sep. 2019.

\bibitem{Li2020BM}
Y.-N.~R. Li, B.~Gao, X.~Zhang, and K.~Huang, ``Beam management in millimeter-wave communications for 5{G} and beyond,'' \emph{IEEE Access}, vol.~8, pp. 13\,282--13\,293, Jan. 2020.

\bibitem{Andrew2021BM}
Y.~Heng, J.~G. Andrews, J.~Mo, V.~Va, A.~Ali, B.~L. Ng, and J.~C. Zhang, ``Six key challenges for beam management in 5.5{G} and 6{G} systems,'' \emph{IEEE Communications Magazine}, vol.~59, no.~7, pp. 74--79, Jul. 2021.

\bibitem{daiELAA2023}
Y.~Lu, Z.~Zhang, and L.~Dai, ``Hierarchical beam training for extremely large-scale {MIMO}: From far-field to near-field,'' \emph{{IEEE} Trans. Commun.}, pp. 1--1, Dec. 2023.

\bibitem{Heath2016OverviewmmWave}
R.~W. {Heath}, N.~{González-Prelcic} \emph{et~al.}, ``An overview of signal processing techniques for {M}illimeter {W}ave {MIMO} systems,'' \emph{{IEEE} J. Sel. Topics Signal Process.}, vol.~10, no.~3, pp. 436--453, Apr. 2016.

\bibitem{attiah2022DLCSHP}
K.~M. Attiah, F.~Sohrabi, and W.~Yu, ``Deep learning for channel sensing and hybrid precoding in {TDD} massive {MIMO} {OFDM} systems,'' \emph{{IEEE} Trans. Commun.}, vol.~21, no.~12, pp. 10\,839--10\,853, Dec. 2022.

\bibitem{Sohrabi2022ActiveSens}
F.~Sohrabi, T.~Jiang, W.~Cui, and W.~Yu, ``Active sensing for communications by learning,'' \emph{{IEEE} J. Sel. Areas Commun.}, vol.~40, no.~6, pp. 1780--1794, Jun. 2022.

\bibitem{Andrews2022BF}
Y.~Heng, J.~Mo, and J.~G. Andrews, ``Learning site-specific probing beams for fast mm{W}ave beam alignment,'' \emph{{IEEE} Trans. Wireless Commun.}, vol.~21, no.~8, pp. 5785--5800, Aug. 2022.

\bibitem{li2019deep}
X.~Li and A.~Alkhateeb, ``Deep learning for direct hybrid precoding in millimeter wave massive {MIMO} systems,'' in \emph{Proc. IEEE Asilomar}, 2019, pp. 800--805.

\bibitem{Khateeb2023GPS}
J.~Morais, A.~Bchboodi, H.~Pezeshki, and A.~Alkhateeb, ``Position-aided beam prediction in the real world: How useful {GPS} locations actually are?'' in \emph{Proc. IEEE Int. Conf. Commun. (ICC)}, Rome, Italy, Jun. 2023, pp. 1824--1829.

\bibitem{khateeb2022radar}
U.~Demirhan and A.~Alkhateeb, ``Radar aided 6{G} beam prediction: Deep learning algorithms and real-world demonstration,'' in \emph{Proc. IEEE Wireless Commun. and Netw. Conf. (WCNC)}, Austin, TX, USA, May 2022, pp. 2655--2660.

\bibitem{Salehi2022}
B.~Salehi, G.~Reus-Muns, D.~Roy, Z.~Wang, T.~Jian, J.~Dy, S.~Ioannidis, and K.~Chowdhury, ``Deep learning on multimodal sensor data at the wireless edge for vehicular network,'' \emph{IEEE Transactions on Vehicular Technology}, vol.~71, no.~7, pp. 7639--7655, Jul. 2022.

\bibitem{Khateeb20203D}
W.~Xu, F.~Gao, S.~Jin, and A.~Alkhateeb, ``3{D} scene-based beam selection for mm{W}ave communications,'' \emph{IEEE Wireless Communications Letters}, vol.~9, no.~11, pp. 1850--1854, Nov. 2020.

\bibitem{dobre2020Hier}
C.~Qi, K.~Chen, O.~A. Dobre, and G.~Y. Li, ``Hierarchical codebook-based multiuser beam training for millimeter wave massive {MIMO},'' \emph{IEEE Transactions on Wireless Communications}, vol.~19, no.~12, pp. 8142--8152, Dec. 2020.

\bibitem{Asmaa2024MADRL}
A.~Abdallah, A.~Celik, M.~M. Mansour, and A.~M. Eltawil, ``Multi-agent deep reinforcement learning for beam codebook design in {RIS}-aided systems,'' \emph{IEEE Transactions on Wireless Communications}, pp. 1--1, Jan. 2024.

\bibitem{Lim2021BT}
S.~H. Lim, S.~Kim, B.~Shim, and J.~W. Choi, ``Deep learning-based beam tracking for millimeter-wave communications under mobility,'' \emph{IEEE Transactions on Communications}, vol.~69, no.~11, pp. 7458--7469, Nov. 2021.

\bibitem{Liu2019statBM}
W.~Liu and Z.~Wang, ``Statistics-assisted beam training for mmwave massive {MIMO} systems,'' \emph{{IEEE} Commun. Lett.}, vol.~23, no.~8, pp. 1401--1404, Aug. 2019.

\bibitem{DRL2023BM}
Q.~Xue, Y.-J. Liu, Y.~Sun, J.~Wang, L.~Yan, G.~Feng, and S.~Ma, ``Beam management in ultra-dense mmwave network via federated reinforcement learning: An intelligent and secure approach,'' \emph{IEEE Trans. on Cognitive Commun. and Netwo.}, vol.~9, no.~1, pp. 185--197, Feb. 2023.

\bibitem{khateeb2022BM}
M.~Alrabeiah, Y.~Zhang, and A.~Alkhateeb, ``Neural networks based beam codebooks: Learning mmwave massive {MIMO} beams that adapt to deployment and hardware,'' \emph{{IEEE} Trans. Commun.}, vol.~70, no.~6, pp. 3818--3833, Jun. 2022.

\bibitem{zhang2021reinforcement}
Y.~Zhang, M.~Alrabeiah, and A.~Alkhateeb, ``Reinforcement learning of beam codebooks in millimeter wave and terahertz {MIMO} systems,'' \emph{{IEEE} Trans. Commun.}, 2022.

\bibitem{AsmaaCCNC2023MADRL}
A.~Abdallah, A.~Celik, M.~M. Mansour, and A.~M. Eltawil, ``Deep reinforcement learning based beamforming codebook design for {RIS}-aided mm{W}ave systems,'' in \emph{Proc. IEEE Consumer Commun. and Netw. Conf. (CCNC)}, Las Vegas, Nevada, USA, Jan. 2023, pp. 1020--1026.

\bibitem{AsmaaGLOBECOM2023MADRL}
------, ``Beamforming codebook design for distributed {RIS} networks using deep reinforcement learning,'' in \emph{Proc. IEEE Global Conmun. Conf. Workshops (GLOBECOM Workshops)}, Kuala Lumpur, Malaysia, Dec. 2023.

\bibitem{heng2023grid}
Y.~Heng and J.~G. Andrews, ``Grid-free {MIMO} beam alignment through site-specific deep learning,'' \emph{{IEEE} Trans. Wireless Commun.}, Feb. 2024.

\bibitem{Hanzo2023BA}
K.~Ma, Z.~Wang, W.~Tian, S.~Chen, and L.~Hanzo, ``Deep learning for mm{W}ave beam-management: State-of-the-art, opportunities and challenges,'' \emph{IEEE Wireless Commun.}, vol.~30, no.~4, pp. 108--114, Aug. 2023.

\bibitem{khan2023BM}
M.~Qurratulain~Khan, A.~Gaber, P.~Schulz, and G.~Fettweis, ``Machine learning for millimeter wave and terahertz beam management: A survey and open challenges,'' \emph{IEEE Access}, vol.~11, pp. 11\,880--11\,902, Feb. 2023.

\bibitem{XAInetworking}
T.~Zhang \emph{et~al.}, ``Interpreting {AI} for networking: Where we are and where we are going,'' \emph{IEEE Commun. Mag.}, vol.~60, no.~2, pp. 25--31, 2022.

\bibitem{khan2024XAImag}
N.~Khan, S.~Coleri, A.~Abdallah, A.~Celik, and A.~M. Eltawil, ``Explainable and robust artificial intelligence for trustworthy resource management in 6{G} networks,'' \emph{IEEE Commun. Mag.}, pp. 1--7, 2024.

\bibitem{AsmaaE2E2024Glob}
A.~Abdallah, A.~Celik, A.~Alkhateeb, and A.~M. Eltawil, ``End-to-end learning of beam probing and {RSSI}-based multi-user hybrid precoding design,'' in \emph{Accepted to appear in IEEE GLOBECOM}, Dec. 2024.

\bibitem{Heath2016shiftOrSwitches}
R.~{Méndez-Rial}, C.~{Rusu}, N.~{González-Prelcic}, A.~{Alkhateeb}, and R.~W. {Heath}, ``Hybrid {MIMO} architectures for millimeter wave communications: Phase shifters or switches?'' \emph{IEEE Access}, vol.~4, pp. 247--267, Jan. 2016.

\bibitem{Poor2020MUBF}
A.~A. Nasir, H.~D. Tuan, T.~Q. Duong, H.~V. Poor, and L.~Hanzo, ``Hybrid beamforming for multi-user millimeter-wave networks,'' \emph{{IEEE} Trans. Veh. Technol.}, vol.~69, no.~3, pp. 2943--2956, Mar. 2020.

\bibitem{khateeb2015MUBF}
A.~Alkhateeb, G.~Leus, and R.~W. Heath, ``Limited feedback hybrid precoding for multi-user millimeter wave systems,'' \emph{{IEEE} Trans. Wireless Commun.}, vol.~14, no.~11, pp. 6481--6494, Nov. 2015.

\bibitem{Trabelsi2017DeepCN}
C.~Trabelsi, O.~Bilaniuk, D.~Serdyuk, S.~Subramanian, J.~F. Santos, S.~Mehri, N.~Rostamzadeh, Y.~Bengio, and C.~J. Pal, ``Deep complex networks,'' \emph{ArXiv}, vol. abs/1705.09792, 2017.

\bibitem{nikbakht2021unsupervised-parametric-optimization}
R.~Nikbakht, A.~Jonsson, and A.~Lozano, ``Unsupervised learning for parametric optimization,'' \emph{{IEEE} Commun. Lett.}, vol.~25, no.~3, pp. 678--681, Sep. 2020.

\bibitem{shwartz2017opening}
R.~Shwartz-Ziv and N.~Tishby, ``Opening the black box of deep neural networks via information,'' \emph{arXiv preprint arXiv:1703.00810}, 2017.

\bibitem{Sanchez2015}
L.~G. Sanchez~Giraldo, M.~Rao, and J.~C. Principe, ``Measures of entropy from data using infinitely divisible kernels,'' \emph{{IEEE} Trans. Inf. Theory}, vol.~61, no.~1, pp. 535--548, Jan. 2015.

\bibitem{theoauto2021}
G.~Boquet, E.~Macias, A.~Morell, J.~Serrano, and J.~L. Vicario, ``Theoretical tuning of the autoencoder bottleneck layer dimension: A mutual information-based algorithm,'' in \emph{Proc. European Signal Processing Conference (EUSIPCO)}, Amsterdam, Netherlands, Jan. 2021, pp. 1512--1516.

\bibitem{yu2019understanding}
S.~Yu and J.~C. Principe, ``Understanding autoencoders with information theoretic concepts,'' \emph{Neural Networks}, vol. 117, pp. 104--123, Sep. 2019.

\bibitem{Alkhateeb2019}
A.~Alkhateeb, ``{DeepMIMO}: A generic deep learning dataset for millimeter wave and massive {MIMO} applications,'' in \emph{Proc. of Info. Theory and Appl. Workshop (ITA)}, San Diego, CA, Feb 2019, pp. 1--8.

\end{thebibliography}

	
\end{document}